\documentclass[manuscript]{aastex}

\newcommand{\difd}[2]{{\partial #1 \over \partial #2}}
\newcommand{\dif}[2]{{\partial #1 / \partial #2}}
\newcommand{\Difd}[2]{{d #1 \over d #2}}
\newcommand{\Dif}[2]{{d #1 / d #2}}
\newcommand{\be}{\begin{equation}}
\newcommand{\ee}{\end{equation}}
\newcommand{\revII}[1]{{#1}} 
\shorttitle{Structure and Mass of Filamentary Cloud}
\shortauthors{Tomisaka}

\begin{document}

\title{Magnetohydrostatic Equilibrium Structure and Mass of Filamentary Isothermal Cloud
Threaded by Lateral Magnetic Field}


\author{Kohji Tomisaka\altaffilmark{1}}
\affil{Division of Theoretical Astronomy, National Astronomical Observatory of Japan,
Mitaka, Tokyo 181-8588, Japan}
\email{tomisaka@th.nao.ac.jp}


\altaffiltext{1}{Department of Astronomical Science, 
School of Physical Sciences,
Graduate University for Advanced Studies (SOKENDAI),
Mitaka, Tokyo 181-8588, Japan}


\begin{abstract}
Herschel observation has recently revealed that
 interstellar molecular clouds consist of many filaments.  
Polarization observations in optical and infrared wavelengths indicate
 that the magnetic field often runs perpendicular to the filament.  
In this paper, the magnetohydrostatic configuration of isothermal gas is studied,
 in which the thermal pressure and the Lorentz force are balanced against
 the self-gravity and the magnetic field is globally perpendicular
 to the axis of the filament. 
The model is controlled by three parameters:  center-to-surface density
 ratio ($\rho_c/\rho_s$),
 plasma $\beta$ of surrounding interstellar gas ($\beta_0$)
 and the radius of the hypothetical parent cloud normalized by the scale-height ($R'_{0}$),
 although there remains a freedom how the mass is distributed
 against the magnetic flux (mass loading).  
In the case that $R'_0$ is small enough,
 the magnetic field plays a role in confining the gas.  
However, the magnetic field generally has an effect in supporting the cloud. 
There is a maximum line-mass (mass per unit length)
 above which the cloud is not supported against the gravity.  
Compared with the maximum line-mass of non-magnetized cloud ($2c_s^2/G$,
 where $c_s$ and $G$ represent respectively the isothermal sound speed
 and the gravitational constant),
 that of the magnetized filament is larger than the non-magnetized one.
The maximum line-mass is numerically obtained as
\[ \lambda_{\rm max} \simeq 0.24 \Phi_{\rm cl}/G^{1/2} + 1.66 c_s^2/G, \]
where $\Phi_{\rm cl}$ represents one half of the magnetic flux threading the filament per unit length.  
The maximum mass of the filamentary cloud is shown to be significantly
 affected by the magnetic field when the magnetic flux per unit length
 exceeds $\Phi_{\rm cl} \gtrsim 3\,{\rm pc\,\mu G}\,(c_s/190\,{\rm m\,s^{-1}})^2$.
\end{abstract}





\section{Introduction}
Recently, filamentary structure in molecular clouds
 has attracted a great deal of attention
 in the context of star formation.
Thanks to the high sensitivity of Herschel satellite
 \citep{pilbratt2010} in infrared (IR) and sub-mm ranges,
\revII{Herschel} has found many filaments in the thermal dust emissions
 from molecular clouds,
 which include clouds inactive in star formation such as Polaris 
 \citep{menshchikov2010,miville2010},
 as well as active ones such as
  the Aquila cloud \citep{menshchikov2010},
  IC 5146 \citep{arzoumanian2011},
  Vela C \citep{hill2011}, and
  Rosette cloud \citep{schneider2012}.
This indicates that the molecular clouds
 consist of the gas filaments
 and the star formation process should be studied
 in this context.

Polarization observation of background stars beyond a molecular cloud  
 gives the geometry of interstellar magnetic field inside and around
 the cloud.
This is based on the fact that the dust is aligned in the magnetic field
 and the light obscured by the intervening aligned interstellar dusts
 residing in the cloud  
 shows a polarization such as the {\bf E}-vector of the polarization
 is parallel to the interstellar magnetic field.
From the near IR (J, H, and Ks bands) imaging polarimetry
 of Serpens South Cloud, 
 \citet{sugitani2011} have found a well-ordered global magnetic field
 perpendicular to the main filament.
They also found that small-scale filaments seem to run along the magnetic field.
Even in the Taurus dark cloud, optical and near IR polarimetry \citep{moneti1984}
 indicates that the global magnetic field seems perpendicular to the major
 axis of the clouds.
B211 and B213 filamentary clouds run in a direction perpendicular    
 to the magnetic field, while many low-density striations seen
 outside of the filament extend along the magnetic
 field \citep{palmeirim2013}.
This geometry is sometimes believed to be an outcome of interstellar
 MHD (magnetohydrodynamic) turbulence \citep{li2006}.  
That is, a turbulent sheet or filament is formed perpendicular to
 the global magnetic field.

There are a number of studies about the filamentary gas cloud
 based on the hydrostatic and magnetohydrostatic equilibria. 
When a filament is sufficiently long compared
 with its width, the cloud can be regarded as an infinitely long cylinder.
\revII{Under the assumptions of axisymmetry and no magnetic field,}
 density distribution of a cylindrical isothermal cloud with
 a central density $\rho_c$ is expressed analytically
 \citep{stodolkiewicz1963,ostriker1964} as
\be
\rho(r)=\rho_c\left(1+\frac{r^2}{8H^2}\right)^{-2} 
\label{eqn:cyl-rho}
\ee
where $H$ is a scale-height and is expressed
 using the isothermal sound speed $c_s$, the central density $\rho_c$,
 and the gravitational constant $G$ as
 $H=c_s/(4\pi G \rho_c)^{1/2}$.
This leads to a mass distribution, which is defined as
 a mass contained inside radius $r$ per unit length, such as
\begin{mathletters}
\begin{eqnarray}
\lambda(r) &=& \int_0^r 2\pi \rho r dr\\
&=&\frac{2c_s^2}{G}\frac{{r^2}/{8H^2}}{1+{r^2}/{8H^2}}.
\label{eqn:cyl-lambda-dens}
\end{eqnarray}
\end{mathletters}
\hspace*{-3mm}
The solution is truncated at a radius $R$ where the pressure $p=\rho c_s^2$
 is balanced with the external ambient pressure $p_{\rm ext}$, that is,
 $\rho \ge p_{\rm ext}/c_s^2$.
Equation (\ref{eqn:cyl-lambda-dens}) shows that
 the cylindrical filament has a maximum line-mass (mass per unit length)
 $\lambda(R) \le \lambda(R=\infty)=2c_s^2/G$,
 which corresponds to the mass of a filament in the vacuum $p_{\rm ext}=0$.
The character of the isothermal filament is controlled by a parameter
 as $\lambda(R)/(2c_s^2/G)$ \citep{nagasawa1987,inutsuka1992,fischera2012}.
Herschel's observations of Aquila and Polaris clouds show that
 the Aquila main cloud has a large line-mass as $\lambda \gtrsim 5 \times (2c_s^2/G)$
 and is rich in protoclusters
 but that a portion with low line-mass as $\lambda \lesssim 2c_s^2/G$
 is devoid of prestellar cores and protostars \citep{andre2010}
 (To derive mass per unit length from observed column density, 
 the width of the filament is assumed constant in
 their paper as 
 FWHM $\sim 14000 {\rm AU}$ in Aquila and $\sim 9000{\rm AU}$ in Polaris).

When the filaments have magnetic fields only parallel to their axes, $B_z$, 
 the structure is also analytically given by equation (\ref{eqn:cyl-rho})
 but in this case $H=c_s(1+\beta^{-1})^{1/2}/(4\pi G \rho_c)^{1/2}$,
 where the plasma beta is assumed to be constant as
 $\beta=c_s^2\rho/(B_z^2/8\pi)={\rm const}$ \citep{stodolkiewicz1963}.  
The mass distribution $\lambda(r)$ increases with the magnetic field
 strength as
\begin{equation}
\lambda(r) = \frac{2c_s^2}{G}\left(1+\beta^{-1}\right)\frac{{r^2}/{8H^2}}{1+{r^2}/{8H^2}}.
\label{eqn:cyl-lambda-mag-dens}
\end{equation}
Comparing equations (\ref{eqn:cyl-lambda-dens}) and 
 (\ref{eqn:cyl-lambda-mag-dens}),
 the line-mass increases in proportion to $1+\beta^{-1}$.
The maximum line-mass supported against the self-gravity (for $r \gg H$)
 increases also in proportion to $1+\beta^{-1}$.
Similar solutions are obtained numerically for the case
 whose mass-to-flux ratio is constant $\Gamma_z\equiv B_z/\rho={\rm const}$ \citep{fiege2000a,fiege2000b}.
In both cases, the poloidal magnetic field $B_z$ has the effect of increasing
 the mass of a static filament.
\citet{fiege2000a,fiege2000b} also consider the effect of toroidal magnetic field $B_\phi$
 assuming a type of flux conservation, $\Gamma_\phi\equiv B_\phi/(r\rho)={\rm const}$. 
In this case, the toroidal magnetic field $B_\phi$ has the opposite effect
 of reducing the supported mass, because the $B_\phi$ exerts the ``hoop stress'' 
 to compress the filament in the radial direction.
\footnote{\revII{The radial Lorentz force coming from $B_\phi$ is proportional
 to the current in $z$-direction, 
 $\propto \dif{rB_\phi}{r}=\dif{r^2\rho\Gamma}{r}$.
 Thus, the direction of the force is determined from the density distribution,
 or $M\equiv \dif{\log\rho}{\log r}+2$.
 \citet{fiege2000a} have shown that $M>0$ for their all isothermal and logatropic
 models, which means this force is working inwardly.}}
However, the relation between the axis of the filament
 and the interstellar magnetic field is observed far from such simple
 configurations previously studied;
 that is, the actual filament is often perpendicular to the global magnetic field
 rather than the filament simply having poloidal and/or toroidal components.
In the present paper,
 we revisit the magnetohydrostatic configuration of isothermal filaments,
 paying attention to the polarization observations, which indicate that
 the global magnetic field is often perpendicular to the filament.

The axisymmetric cloud threaded by the poloidal magnetic field has
 a similar maximum mass that depends on the magnetic flux.
\revII{From numerically obtained magnetohydrostatic configurations,
 it is shown that
 the maximum column density $\sigma_{\rm cr}$ depends on the magnetic flux density $B_0$ as
 $\sigma_{\rm cr}\sim 0.17 B_0/G^{1/2}$(eq.(4.8) of \citet{tomisaka1988b}).}
This gives the maximum mass $M_{\rm cr}$ is proportional to the magnetic flux $\phi_0$,
 $M_{\rm cr}\sim 0.17 \phi_0/G^{1/2}$.
\revII{This maximum column density is nearly equal to the maximum stable column density
 against the gravitational instability
 of magnetized plane-parallel sheet, $\sigma_{\rm cr} = B_0/(2\pi G^{1/2})$ obtained
 by \citet{nakano78}.}

The filamentary structure may be in dynamical contraction \citep{inutsuka1992, kawachi1998}
 rather than in a hydrostatic state considered here.
However the condition to begin the dynamical contraction is given from
 the hydrostatic maximum line-mass supported against the self-gravity. 

The structure of this paper is as follows:
 in $\S$2, the model and formulation for obtaining a magnetohydrostatic
 configuration are given.
The method is a self-consistent field method
 similar to \citet{mouschovias1976a,mouschovias1976b} and
 \citet{tomisaka1988a,tomisaka1988b}, although these authors considered a disk-like cloud
 threaded perpendicularly by the magnetic field.
Formulation for the filament with a lateral magnetic field is described  
 in the following section.
We show the numerical result in $\S$ 3,
 in which structure of the filament is shown.
Discussion on the effect of a magnetic field,
 such as how much mass is supported
 by the lateral magnetic field,
 is shown in $\S$ 4. 
Section 5 is devoted to a summary.

\section{Method and Model}
\subsection{Magnetohydrostatic Equations}
Basic equations to obtain magnetohydrostatic configurations
 of isothermal gas are composed of three equations:
 the force balance equation between the Lorentz force, gravity and the pressure
 force, the Poisson equation for gravitational potential $\psi$,
 and the Ampere's law between the current {\bf j} and the magnetic flux density
 {\bf B} as follows: 
\be
\frac{1}{c}{\bf j}\times {\bf B} - \rho \nabla \psi - c_s^2 \nabla \rho =0,
 \label{eqn:force-balance}
\ee
\be
\nabla^2 \psi = 4 \pi G \rho,
 \label{eqn:poisson-eq}
\ee
\be
{\bf j}=\frac{c}{4\pi} \nabla \times {\bf B},
 \label{eqn:ampere-law}
\ee
where $\rho$, $c_s$, $c$, and $G$ represent, respectively,
 the gas density, isothermal sound speed, speed of light
 and Newton's constant of gravity.
We assume here a filament is extending along the $z$-axis and assume the 
 filament is also uniform in the $z$-direction in the Cartesian coordinate
 $(x, y, z)$.
We use a flux function $\Phi(x,y)$ to calculate the magnetic flux density
 ${\bf B}$ as
\begin{mathletters}
\begin{eqnarray}
B_x&=&-\difd{\Phi}{y},\\
B_y&=&\difd{\Phi}{x}.
\end{eqnarray}
\end{mathletters}
Although we will call this flux function $\Phi$
 as a magnetic flux of a cylindrical cloud in the present paper,
 $\Phi$ has a dimension of the magnetic flux density $B$ times the size
 $L$, that is, $[\Phi]=[B][L]$ not an ordinary magnetic flux $[B][L]^2$.
Since $\dif{}{z}=0$, $\Phi$ is the $z$-component of the natural vector potential,
 $\Phi=A_z$.
Assuming $\dif{}{z}=0$, from the Ampere's law (eq.[\ref{eqn:ampere-law}])
 the electric current is given as 
\begin{mathletters}
\begin{eqnarray}
j_x&=&\frac{c}{4\pi}\difd{B_z}{y},\\
j_y&=&-\frac{c}{4\pi}\difd{B_z}{x},\\
j_z&=&\frac{c}{4\pi}\left(\difd{B_y}{x}-\difd{B_x}{y}\right)=
 -\frac{c}{4\pi}\Delta_2\Phi,
\end{eqnarray}
\end{mathletters}
where $\Delta_2 \equiv \partial^2/\partial x^2+\partial^2/\partial y^2$.
The $z$-component of equation (\ref{eqn:force-balance}) is reduced to
${\bf j}\times {\bf B}|_z=0$ (``force-free'' condition) and thus
\be
\left(\difd{B_z}{y}\right)\left(\difd{\Phi}{x}\right)
-\left(\difd{B_z}{x}\right)\left(\difd{\Phi}{y}\right)=0.
\ee
Since this equation is rewritten as
\be
\frac{\left(\dif{B_z}{y}\right)}{\left(\dif{\Phi}{y}\right)}
=\frac{\left(\dif{B_z}{x}\right)}{\left(\dif{\Phi}{x}\right)},
\ee
this requires that $B_z$ depends only on the flux function as
 $B_z=B_z(\Phi)$.
The $x$- and $y$-components of equation (\ref{eqn:force-balance}) are reduced to
\begin{mathletters}
\begin{eqnarray}
\frac{1}{4\pi}\Delta_2\Phi\difd{\Phi}{x}
 -\rho\difd{\psi}{x}-c_s^2\difd{\rho}{x}-\frac{1}{8\pi}\difd{B_z^2}{x}&=&0,\\
\frac{1}{4\pi}\Delta_2\Phi\difd{\Phi}{y}
 -\rho\difd{\psi}{y}-c_s^2\difd{\rho}{y}-\frac{1}{8\pi}\difd{B_z^2}{y}&=&0,
\end{eqnarray}
\end{mathletters}
in which the last term of the left-hand side represents the magnetic pressure force.
In this paper, we restrict ourselves to the $B_z=0$ model.
In this case, the force balance is simply reduced to the following equations:
\begin{mathletters}
\begin{eqnarray}
\frac{1}{4\pi}\Delta_2\Phi\difd{\Phi}{x}
 &=&\rho\difd{\psi}{x}+c_s^2\difd{\rho}{x},\label{eqn:comp-force-balancea}
\\
\frac{1}{4\pi}\Delta_2\Phi\difd{\Phi}{y}
 &=&\rho\difd{\psi}{y}+c_s^2\difd{\rho}{y}.\label{eqn:comp-force-balanceb}
\end{eqnarray}
\end{mathletters}
Since the Lorentz force exerts no force in the direction of the magnetic field, 
 the force balance along the magnetic field, i.e., in the direction of 
 $(B_x,B_y)$ is expressed as 
\be
-\rho\difd{\psi}{s}-c_s^2\difd{\rho}{s}=0,
\ee
where $s$ represents the distance measured along the magnetic field line.
Integrating along the magnetic field line, the density is expressed as 
\be
\rho=\frac{q}{c_s^2}\exp{\left(-\frac{\psi}{c_s^2}\right)},\label{eqn:rho-s}
\ee  
where $q$ is an integration constant determined for each magnetic field line
 and thus $q(\Phi)$ is a function of $\Phi$.
Using equation (\ref{eqn:rho-s}),
 the right-hand side of equations (\ref{eqn:comp-force-balancea}) and
(\ref{eqn:comp-force-balanceb}) become respectively
$(\dif{q}{x})\exp{\left(-\psi/c_s^2\right)}$ and
$(\dif{q}{y})\exp{\left(-\psi/c_s^2\right)}$.
Considering the fact that $q$ is a function of $\Phi$,
 equations (\ref{eqn:comp-force-balancea}) and (\ref{eqn:comp-force-balanceb}) 
 require 
\be
\Delta_2\Phi = 4 \pi \Difd{q}{\Phi}\exp{\left(-\frac{\psi}{c_s^2}\right)}.
\label{eqn:poisson-Phi}
\ee
The other equation to be solved is the Poisson equation for the gravitational
 potential $\psi$ (eq.[\ref{eqn:poisson-eq}]) as
\be
\Delta_2\psi = 4 \pi G \frac{q(\Phi)}{c_s^2}\exp{\left(-\frac{\psi}{c_s^2}\right)}.
\label{eqn:poisson-psi}
\ee
Equations (\ref{eqn:poisson-Phi}) and (\ref{eqn:poisson-psi}) are
 the basic equations,
 which are a coupled partial differential equation system
 of the elliptic type for the two variables
 $\Phi$ and $\psi$ after $q(\Phi)$ is determined. 
We search for the solutions of $\psi$ and $\Phi$ that
 simultaneously satisfy equations (\ref{eqn:poisson-Phi}) and (\ref{eqn:poisson-psi}) 
 by the self-consistent field method.
We assume initial guesses for $\psi$ and $\Phi$ and let them converge to 
 true solutions.
 
\subsection{Mass Loading}
\begin{figure}
\plotone{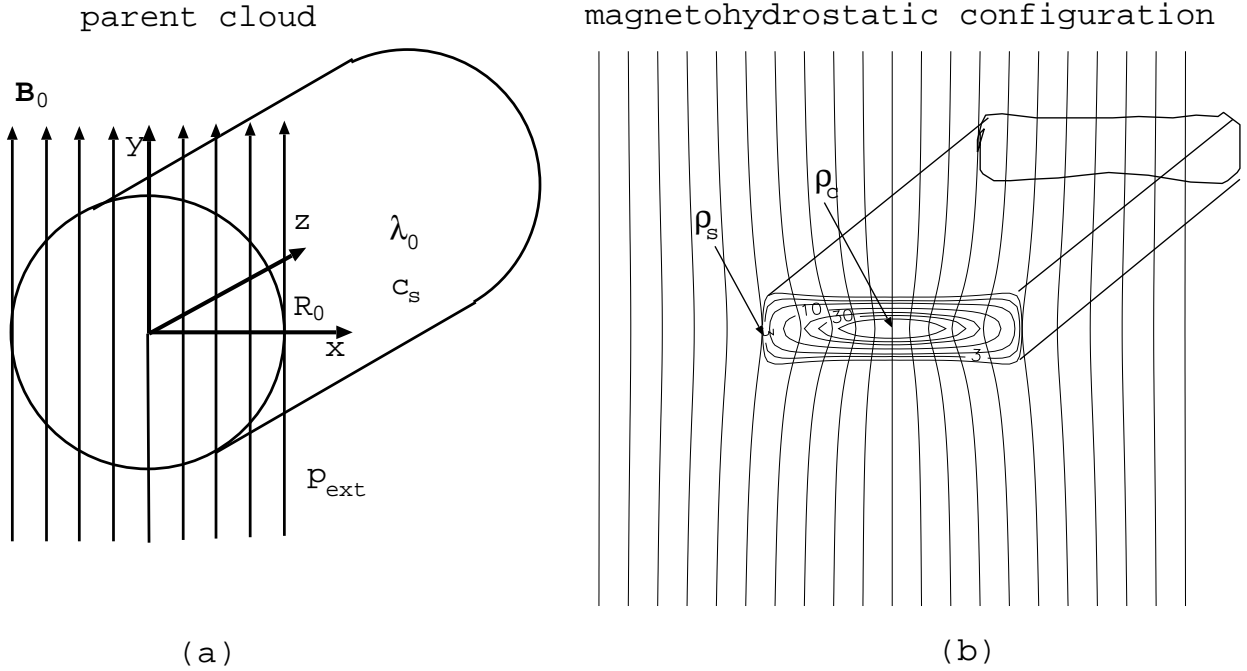}
\caption{Explanation about model parameters.
In ({\it a}), we show the parent cloud with radius $R_0$,
 which has a uniform density and is threaded by a uniform magnetic field
 with the strength of $B_0$.
The line-mass (mass per unit length) of the filamentary cloud
 is equal to $\lambda_0$.
The cloud is immersed in an ambient pressure $p_{\rm ext}$.
The cloud extends in the $z$-direction and the magnetic field runs
 in the $y$-direction.
Keeping the mass distribution against the magnetic flux (flux freezing)
 of the parent cloud ({\it a}), we search for a magnetohydrostatic 
 configuration shown in ({\it b}).
Since the cloud must satisfy the pressure equilibrium at the surface,
 the density at the surface of the equilibrium state ({\it b}) is equal to
 $\rho_s=p_{\rm ext}/c_s^2$, where $c_s$ represents the isothermal sound speed
 in the cloud.
The model is specified by the normalized radius $R_0'\equiv
 R_0/[c_s/(4\pi G \rho_s)^{1/2}]$, 
 plasma $\beta$ of the ambient gas $\beta_0\equiv p_{\rm ext}/(B_0^2/8\pi)$,
 and the normalized line-mass
 as $\lambda'_0\equiv \lambda_0/(c_s^2/4 \pi G)$.
It is more convenient to use
 the normalized central density $\rho'_c\equiv \rho_c/\rho_s$
 rather than the normalized line-mass.
\label{fig1}}
\end{figure}

The function $q(\Phi)$ is calculated based on 
 the line-mass distribution against the magnetic flux per unit length,
 which is sometimes called mass loading.
The line-mass $\Delta \lambda$ between two field lines specified
 by $\Phi$ and $\Phi+\Delta \Phi$ is calculated to first order in $\Delta\Phi$ as
\begin{eqnarray}
\Delta \lambda(\Phi)
&=&2\int_0^{y_s(\Phi)}dy\int_{x(y,\Phi)}^{x(y,\Phi+\Delta \Phi)}dx \rho(x,y)\nonumber \\
&=&2\int_0^{y_s(\Phi)}dy\frac{\rho}{(\dif{\Phi}{x})}\Delta \Phi\nonumber \\
&=&2\int_0^{y_s(\Phi)}dy\frac{q(\Phi)}{c_s^2}\frac{\exp{(-\psi/c_s^2)}}{(\dif{\Phi}{x})}\Delta \Phi,
\end{eqnarray}
where $y_s(\Phi) > 0$ is the $y$-coordinate of the surface of the cloud
 where the density is equal to the density at the surface
 $\rho(y_s(\Phi))=\rho_s\equiv p_{\rm ext}/c_s^2$ (see Fig.\ref{fig1}b).
(In the present paper, we assume that all the physical quantities have
 mirror symmetries against the $x$- and $y$-axes.)  
Thus, the mass-to-flux ratio is calculated as
\be
\Difd{\lambda}{\Phi}=\frac{2q(\Phi)}{c_s^2}\int_0^{y_s(\Phi)}
 \frac{\exp{(-\psi/c_s^2)}}{(\dif{\Phi}{x})}dy,
\label{eqn:dlambda_dPhi}
\ee
where the integration of the right-hand side of the equation
 can be evaluated for the approximate solutions of $\psi$ and $\Phi$
 even if they are not converged yet.

Consider a cylindrical cloud (parent cloud) with a uniform density $\rho_0$ 
 and a radius $R_0$
 which is threaded by a uniform magnetic field $B_0$ (see Fig.\ref{fig1}).
Line mass $\Delta \lambda$ contained between two magnetic field lines specified by 
 $\Phi$ and $\Phi+\Delta \Phi$ is 
\be
\Delta \lambda= 2 \left(R_0 \frac{\Delta \Phi}{\Phi_{\rm cl}}\right)
 \left(R_0\left[1-(\Phi/\Phi_{\rm cl})^2\right]^{1/2}\right) \rho_0,
\ee
where $\Phi_{\rm cl}$ is a total flux per unit length of a cloud defined as 
\be
\Phi_{\rm cl}=R_0B_0,
\ee
 and the flux function
 $\Phi$ varies from $-\Phi_{\rm cl}$ to $+\Phi_{\rm cl}$.
Thus, the mass-to-flux distribution for this uniform cylinder is written as
\be
\Difd{\lambda}{\Phi}=2R_0^2\rho_0 \frac{\left[1-(\Phi/\Phi_{\rm cl})^2\right]^{1/2}}
{\Phi_{\rm cl}}\ \ \ \ (-\Phi_{\rm cl}\le\Phi\le\Phi_{\rm cl}).
\label{eqn:model_dm_dPhi}
\ee 
Using the total line mass
 $\lambda_0=2\int_0^{\Phi_{\rm cl}}(\Dif{\lambda}{\Phi})d\Phi=\pi R_0^2 \rho_0$,
 equation (\ref{eqn:model_dm_dPhi}) is rewritten as
\be
\Difd{\lambda}{\Phi}=\frac{2\lambda_0}{\pi \Phi_{\rm cl}}
 \left[1-(\Phi/\Phi_{\rm cl})^2\right]^{1/2}.
\label{eqn:dlambda_dPhi0}
\ee
If we require that the line-mass distribution of the solution
 ($d\lambda/d\Phi$ of eq.[\ref{eqn:dlambda_dPhi}]) should be equal to
 that of the uniform cylinder (eq.[\ref{eqn:dlambda_dPhi0}]),
 $q(\Phi)$ is calculated as follows:
\be
q(\Phi)=\frac{c_s^2\lambda_0\left[1-(\Phi/\Phi_{\rm cl})^2\right]^{1/2}}{\pi\Phi_{\rm cl}\int_0^{y_s(\Phi)}\exp(-\psi/c_s^2)/(\dif{\Phi}{x})dy},
\label{eqn:obtain_q}
\ee
which can be coupled with the basic equations (\ref{eqn:poisson-Phi}) and
 (\ref{eqn:poisson-psi}).
These three equations are sufficient to describe the magnetohydrostatic 
 configuration.
\revII{The structure and the line-mass are affected by mass loading of the filament.
Difference depending on the mass-loading will be quantitatively discussed 
in a forthcoming paper.}

We normalize the basic equations using 
 the surface density $\rho_s=p_{\rm ext}/c_s^2$,
 isothermal sound speed $c_s$,
 free-fall time $(4\pi G \rho_s)^{-1/2}$,
 scale-height $H=c_s/(4\pi G \rho_s)^{1/2}$,
 and unit magnetic strength $B_u=(8\pi c_s^2\rho_s)^{1/2}$.
Dependent and independent variables are normalized as follows:
 $\rho=\rho' \rho_s$,
 $\psi=\psi' c_s^2$,
 $\lambda=\lambda' \rho_s H^2$,
 $\Phi=\Phi' H B_u$,
 $q=q' \rho_s c_s^2$,
 and ${\bf r}=H {\bf r}'$,
 where the variables with $'$ represent the normalized ones.
Equations (\ref{eqn:poisson-Phi}) and (\ref{eqn:poisson-psi}) reduce respectively
to
\begin{eqnarray}
\Delta_2'\Phi'& = &-\frac{1}{2}\Difd{q'}{\Phi'}\exp{\left(-\psi'\right)}.
\label{eqn:final_basic_equation_1}\\
\Delta_2'\psi'& = & q'\exp{\left(-\psi'\right)}.
\label{eqn:final_basic_equation_2}
\end{eqnarray}
Considering the magnetic field line running along the $y$-axis,
 the central density $\rho_c$ is written in terms of the central gravitational potential
 $\psi_c$ and the central $q(\Phi=0)=q_c$ as
\be
\rho_c=\frac{q_c}{c_s^2}\exp{(-\psi_c/c_s^2)},
\ee
and thus the central $q'_c$ is expressed as
\be
q'_c=\rho_c'\exp{\psi'_c}.
\label{eqn:q'}
\ee
Using this, equation (\ref{eqn:dlambda_dPhi}) gives the mass-to-flux ratio
 of the central flux tube of $\Phi'=0$ and $x'=0$ as
\be
\left. \Difd{\lambda'}{\Phi'}\right|_c=2q'_c \int_0^{y'_s(\Phi'=0)}
\left[\exp(-\psi')/\left(\dif{\Phi'}{x'}\right)\right]_{x'=0}dy'.
\label{eqn:dlambda_from_q'c}
\ee
Equation (\ref{eqn:model_dm_dPhi}) gives the mass-to-flux ratio of $\Phi'\ne 0$
 from that of $\Phi'=0$ as
\be
\Difd{\lambda'}{\Phi'}=\left. \Difd{\lambda'}{\Phi'}\right|_c
 \left[1-\left(\frac{\Phi'}{\Phi'_{cl}}\right)^2\right]^{1/2}.
\label{eqn:dm_dPhi_from_central_value}
\ee
Finally, from equation (\ref{eqn:obtain_q}) $q'$ for $\Phi' \ne 0$ is obtained
\be
q'=\frac{\Dif{\lambda'}{\Phi'}}{2\int_0^{y'_s(\Phi')}\exp{(-\psi')}/(\dif{\Phi'}{x'})dy'},
\label{eqn:q'2}
\ee 
where $\Dif{\lambda'}{\Phi'}$ is calculated
 from equation (\ref{eqn:dm_dPhi_from_central_value})
 using equations (\ref{eqn:q'}) and (\ref{eqn:dlambda_from_q'c}).
This equation gives $q'(\Phi')$ as a function of $\rho_c$,
 although equation (\ref{eqn:obtain_q}) needs to specify $\lambda_0$.
Equations (\ref{eqn:final_basic_equation_1}) and 
 (\ref{eqn:final_basic_equation_2}) with this equation
 to derive $q'$ are the basic equations and
 we find solutions to simultaneously satisfy these two partial differential equations
 using the self-consistent field method. 

The outer boundary condition for these partial differential equations
 of elliptic type is set by
\begin{eqnarray}
\psi&=&2G\lambda_0\log{r}+C,\\
\Phi&=&B_0 x,
\end{eqnarray}
where $r=(x^2+y^2)^{1/2}$ represents the distance from the center of the filament,
 the former condition represents the gravitational potential far
 from the filament being approximated by that of an infinitesimally thin filament with the
 same total line-mass $\lambda_0$ and the latter means the magnetic field
 is uniform with $B_0$ and running in the $y$-direction far from the filament.
These are reduced to non-dimensional form:
\begin{eqnarray}
\psi'&=&\frac{\lambda_0'}{2\pi}\log{r'}+C',\\
\Phi'&=&\beta_0^{-1/2}x',
\end{eqnarray}
where $\beta_0$ represents the plasma $\beta$ outside and far from the filamentary cloud as
 $\beta_0=p_{\rm ext}/(B_0^2/8\pi)$.

The model is specified with three non-dimensional parameters:
 $\beta_0$: plasma beta outside the filament,
 $R_0'$: the radius of the parent filamentary cloud,
 and $\lambda_0'$: total line mass.
Since it is very convenient to choose the central density
 $\rho_c'=\rho_c/\rho_s$ as the last parameter rather than the total mass,
 we use not $\lambda_0'$ but $\rho_c'$ to specify the model.
See Figure \ref{fig1} for explanation. 
The reason for this substitution comes from the fact
 that $\lambda_0'$ has a maximum value above which no equilibrium solution exists
 while $\rho'_c$ does not have such an upper limit
 and the maximum of $\lambda_0'$ cannot be known a priori.
To change the last parameter from $\lambda'_0$ to $\rho_c'$,
 equation (\ref{eqn:q'2}) is derived from equation (\ref{eqn:obtain_q})
 and equation (\ref{eqn:q'2}) enables determination of $q'$ as a function
 of $\rho'_c$ not $\lambda'_0$. 
In our calculation, the total line-mass $\lambda_0'$ is obtained
 after the static configuration is calculated.
For simplicity,
 we will hereafter abbreviate the prime representing the normalized quantity,
 unless the meaning of the quantity is unclear. 
Model parameters are summarized in Table \ref{table1}.
Although we calculated 12 cases of the central density for each model as
 $\rho_c=2$, 3, 5, 10, 20, 30, 50, 100, 200, 300, 500, and $10^3$,
 it was difficult to obtain solutions especially for the models with high central density.
We encountered an oscillation rather than smooth convergence in the converging
 scheme to find a solution.
This oscillation is seen in radially outermost regions of 
 geometrically thin (flat) filaments, which appear in the high central density 
 models.   
In Table \ref{table1}, the range of the central density is shown in which 
 the self-consistent solution is obtained in the last column.
To solve the two-dimensional Poisson equation, we applied the finite-difference
 method to the Laplacian $\Delta_2$ in equations (\ref{eqn:final_basic_equation_1})
 and (\ref{eqn:final_basic_equation_2}).
The number of finite-difference cells
 is 641$\times$641 and the 321th cells are located on the $x$- and $y$-axes.
Compared with a low-resolution study of 161$\times$161,
 the obtained line-mass differs only 0.2\% for typical models ($\rho_c=50$ of Model C3). 
Thus, the numerical convergence is sufficient. 
ICCG (Incomplete Cholesky factorization Conjugate Gradient) algorithm \citep{barrett1994} is used
 to solve the Poisson equation.

\section{Result}

\subsection{Models with Small $R_0$}
\begin{figure}
\epsscale{1.0}
\hspace*{2cm}(a)\hspace*{7cm}(b)\\
\plottwo{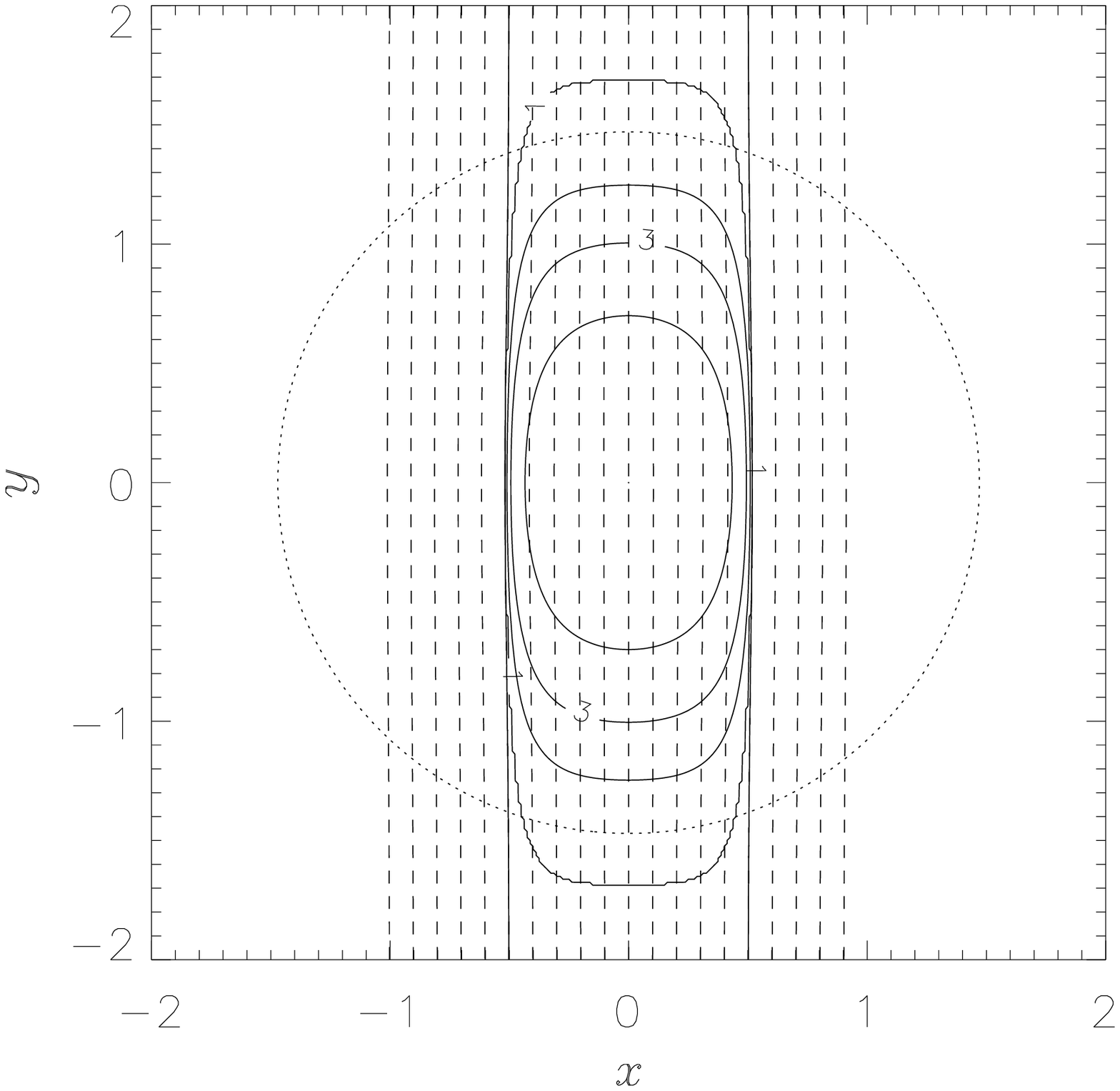}{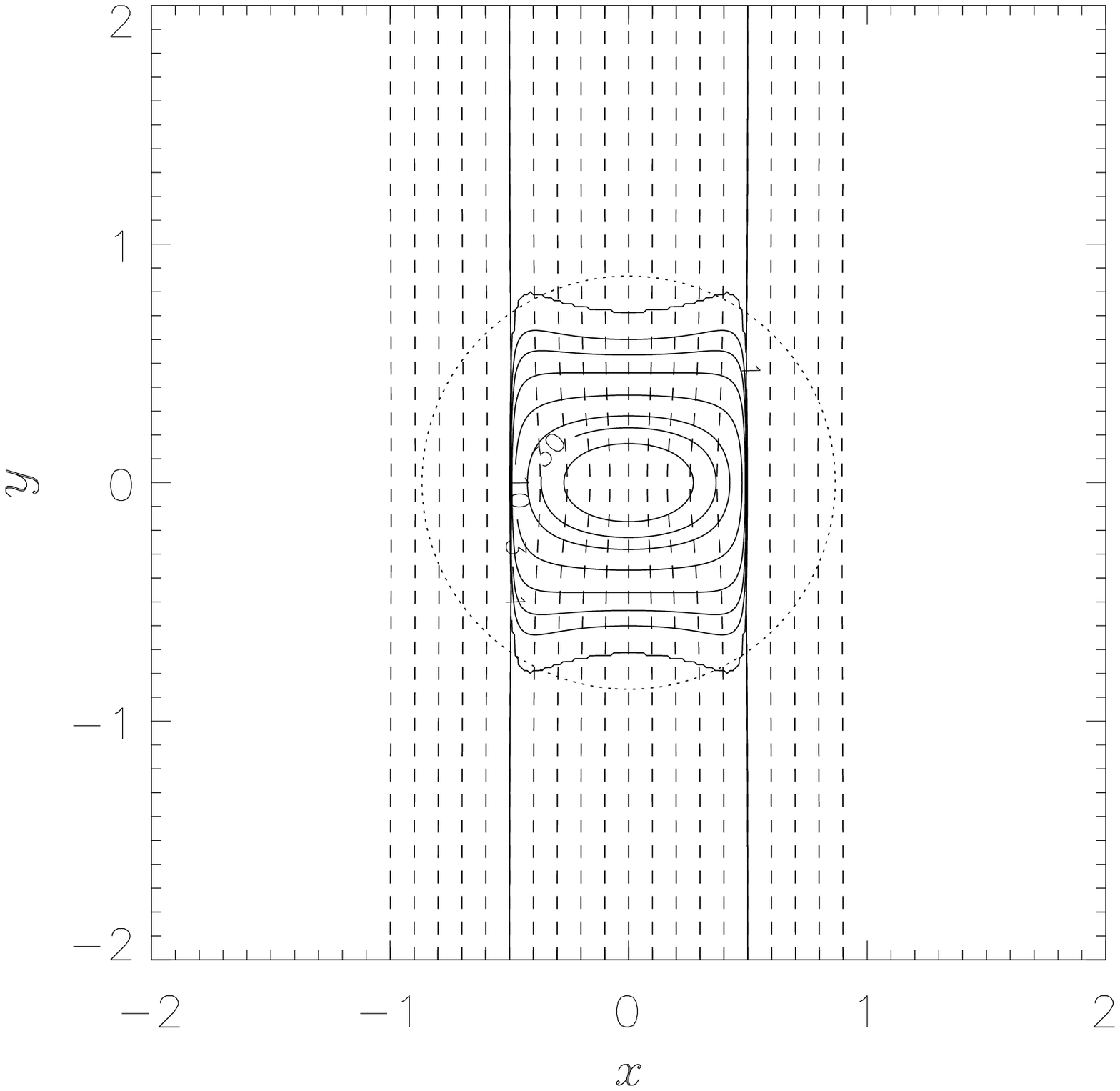}\\
\epsscale{0.5}
\hspace*{4cm}(c)\\
\plotone{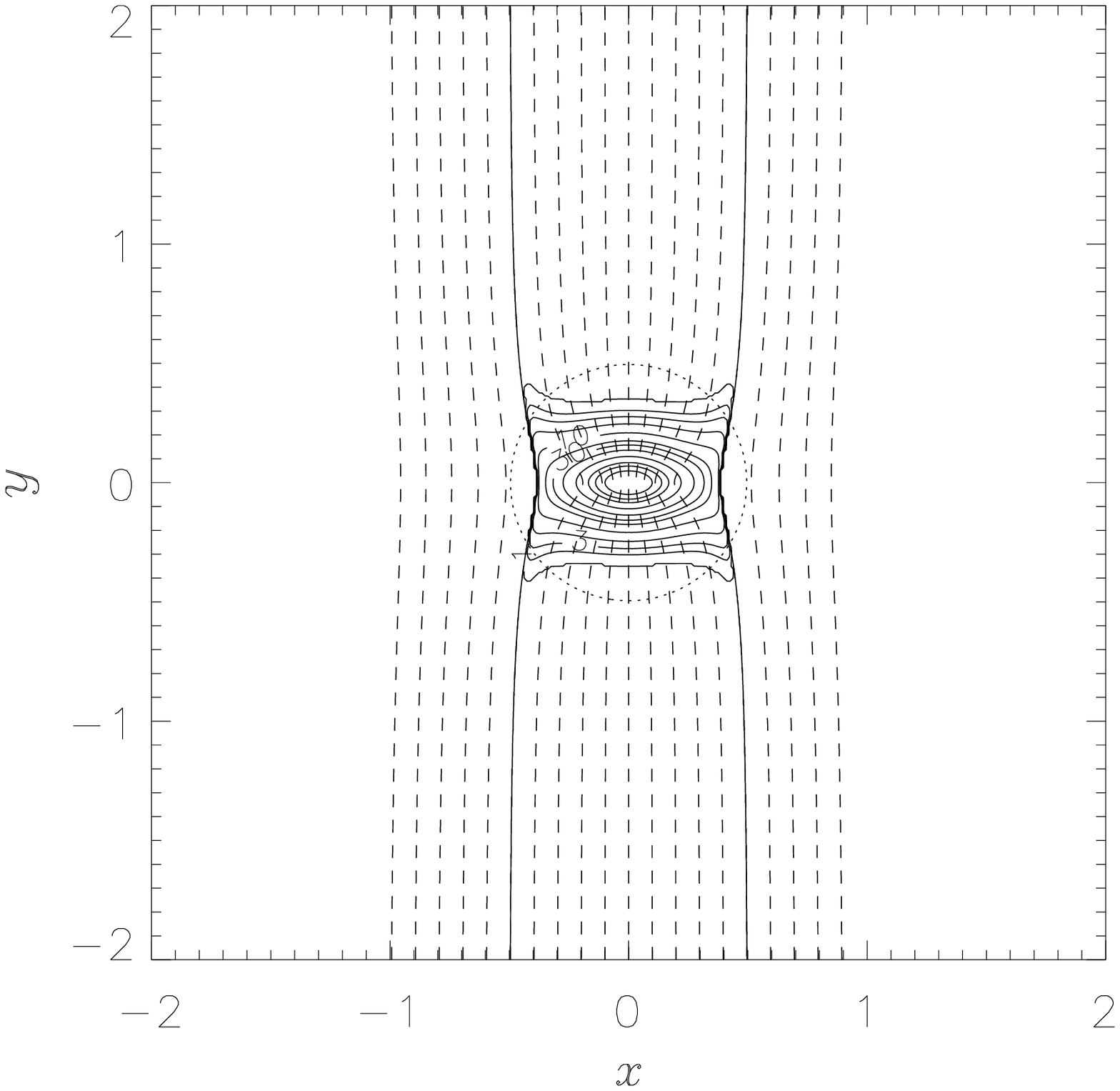}
\caption{Structure of Model A ($R_{\rm cl}=0.5$ and $\beta_0=0.03$).
Each panel corresponds to different central densities
 as $\rho_c=10$ ({\it a}), $\rho_c=100$ ({\it b}), and $\rho_c=10^3$ ({\it c}).
Dashed lines running vertically represent magnetic field lines
 and solid open lines also represent the magnetic field lines but
 specifically those that touch the cloud boundary.
Solid closed lines are the isodensity contours and the contour levels
 are chosen as 1, 2, 3, 5, 10, 20, 30, 50, 100, 200, 300, and 1000.
Dotted circle corresponds to the radius of non-magnetized isothermal
 filament with the same $\rho_c$. 
Models have a line-mass of $\lambda_0=12.9$ ({\it a}),
 $24.9$ ({\it b}),
 and $31.6$ ({\it c}), respectively.
\label{fig2}
}
\end{figure}

We begin with the models with small $R_0 \lesssim 1$, Models A and B.
The model parameters, $R_0$, $\beta_0$, and the range of $\rho_c/\rho_s$ are summarized
 in Table \ref{table1}.
Model A assumes small $R_0=0.5$ and strong magnetic field  $\beta_0=0.03$.
Figure \ref{fig2} illustrates the structures of three states of Model A with different 
 central densities ({\it a}: $\rho_c=10$, {\it b}: $\rho_c=100$, and {\it c}: $\rho_c=10^3$).
As is clearly shown, 
 the vertical size is larger than the horizontal one
 in the models with the low central density as $\rho_c \lesssim 10^2$ 
 (Figs. \ref{fig2} [{\it a}] and [{\it b}]).
On the other hand,
 in the model with high central density (Fig.~2[{\it c}])
 the vertical size is smaller,
 which is ordinarily seen for the magnetized cloud.

Although the magnetic field line seems straight, the shape 
 of the field line differs between the \revII{model} with low central density ({\it a})
 and that with high central density ({\it c}).
The \revII{model} with $\rho_c=10$ (Fig.\ref{fig2}[{\it a}]) has
 magnetic field lines which bow outwardly.
That is, $(B_x,B_y)=(+,+)$ in the fourth quadrant while $(B_x,B_y)=(-,+)$
 in the first quadrant.
In this case, the Lorentz force is exerted radially inwardly,
 which contributes to the shape of the filament in which the major axis
 coincides with the direction of the magnetic field.
On the other hand,
 in the \revII{model} with $\rho_c=10^3$ (Fig.\ref{fig2}[{\it c}]), the 
 magnetic field lines bow inwardly ($(B_x,B_y)=(-,+)$ in the fourth 
 quadrant while $(B_x,B_y)=(+,+)$ in the first quadrant). Thus,
 the Lorentz force is toward outward direction.
Since the Lorentz force has no component in the direction of the magnetic field (vertically), 
 the filament preferentially contracts in the vertical direction.
As a result the major axis of the gas distribution is perpendicular to the direction of
 the magnetic field for models with high central densities.

When axisymmetric clouds (not filaments) are considered,
 a prolate shape that extends in the direction of the magnetic field
 is expected either in a cloud which is overlapped by a toroidal magnetic field
 and is pinched by the magnetic hoop stress \citep{tomisaka1991,fiege2000c} or in a cloud which has a small $R_0$ and
 the magnetic field plays a role not in supporting but in confining the cloud \citep{caitaam2010}.
In the geometry of the filament,
 the magnetic field plays a similar role in the models 
 with small initial radius $R_0$.

The relation between the mass and the central density indicates the stability
of the cloud \revII{(for spherically-symmetric polytropes, see \citet{bonnor1958}; for magnetized clouds, see \citet{tomisaka1988b})}.
That is, when a filament has a line-mass
 that exceeds the maximum allowable one from the relation between the mass
 and the central density,
 the filament has no magnetohydrostatic
 equilibrium and it must undergo a dynamical contraction.
Shown for the disk-like cloud, when one mass corresponds to two central densities,
 one of the two is stable and the other is unstable (\citet{zeldovich1971};
 for the specific case of isothermal magnetized clouds,
 see \S IVb of \citet{tomisaka1988b}). 
We plot the line-mass $\lambda_0$ against the central density $\rho_c$
 for the filamentary cloud.

\begin{figure}
\epsscale{1.0}
\hspace*{2cm}(a)\hspace*{7cm}(b)\\
\plottwo{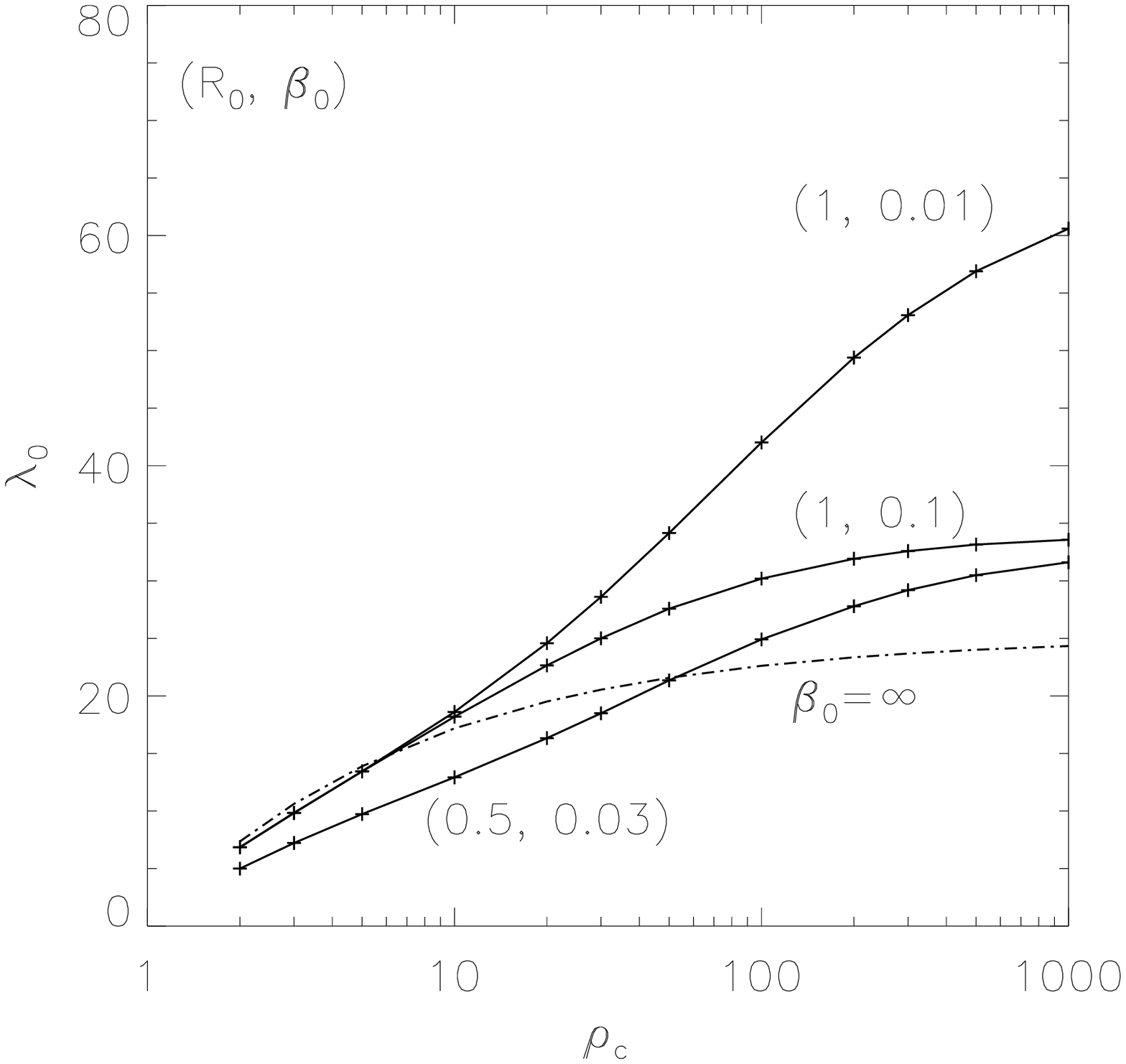}{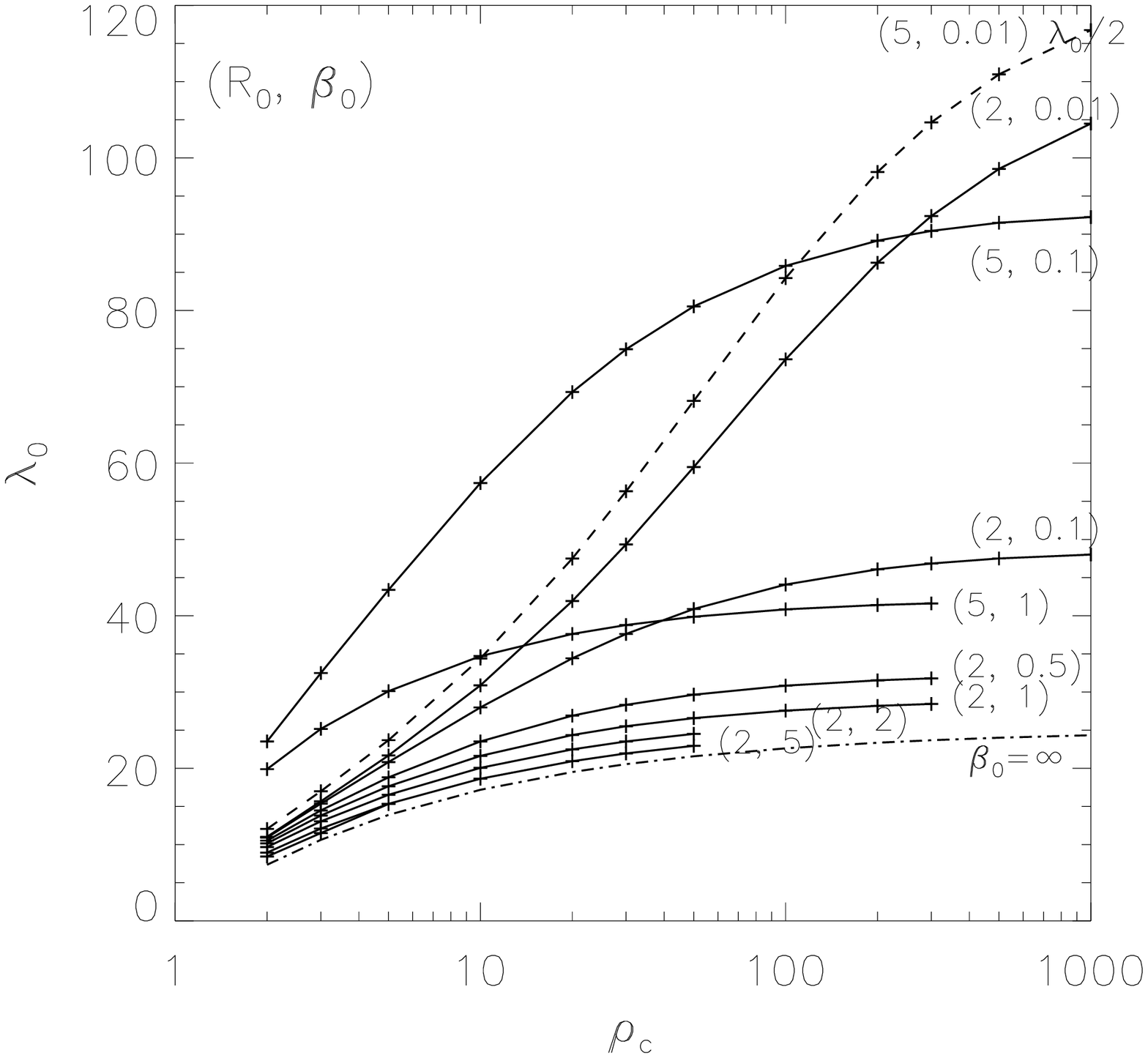}
\caption{The line-mass $\lambda_0$ is plotted against the central density $\rho_c$.
In ({\it a}), models with small $R_{\rm 0}$ (Models A and B) are shown.
The relation of the non-magnetized filament is also shown with the dash-dotted line
 ($\beta_0=\infty$).  
The mass of the filament is smaller than that of the non-magnetized filament 
 for model A with $\rho_c\lesssim 50$.
Even Models B1 and B2, the filament with a low central density $\rho_c\lesssim 5$ is
 less massive than the non-magnetized one (the dash-dotted line).
In ({\it b}), models with larger $R_0=2$ and 5 are plotted.
These models are more massive than the non-magnetized filament.
Only for Model D3 ($R_0=5$ and $\beta_0=0.01$),
 $\lambda_0/2$ is plotted instead of $\lambda_0$ in the dashed line.   
\label{fig3}}
\end{figure}

From equations (\ref{eqn:cyl-rho}) and (\ref{eqn:cyl-lambda-dens}),
 the line-mass of the non-magnetized filament is written down
 by the normalized central density $\rho'_c\equiv \rho_c/\rho_s$ as
\be
\lambda_{\rm 0}=\frac{2c_s^2}{G}\left(1-{\rho'_c}^{-1/2}\right),
\label{eqn:lambda-dim}
\ee
which is rewritten in the non-dimensional form as
\be  
\lambda'_{\rm 0}=8\pi\left(1-{\rho'_c}^{-1/2}\right).\label{eqn:nonmag-lambda}
\label{eqn:lambda-nondim}
\ee
Figure \ref{fig3} ({\it a}) illustrates the line-mass against the normalized central density
 for Models A and B.
Compared with the dash-dotted line, which represents equation (\ref{eqn:nonmag-lambda}),
 it is shown that Model A with $\rho_c\lesssim 50$ is less massive than the non-magnetized filament. 
Figure \ref{fig3} ({\it a}) also indicates that even the filament of Model B
 is less massive than the non-magnetized one when $\rho_c \lesssim 5$.  
This means that the magnetic field plays a role in confining the filament
 in the models with small $R_0$ especially for low central density.
In this case, the magnetic field has a role in the reduction of the equilibrium mass.

\begin{figure}
\epsscale{1.0}
\hspace*{2cm}(a)\hspace*{7cm}(b)\\
\plottwo{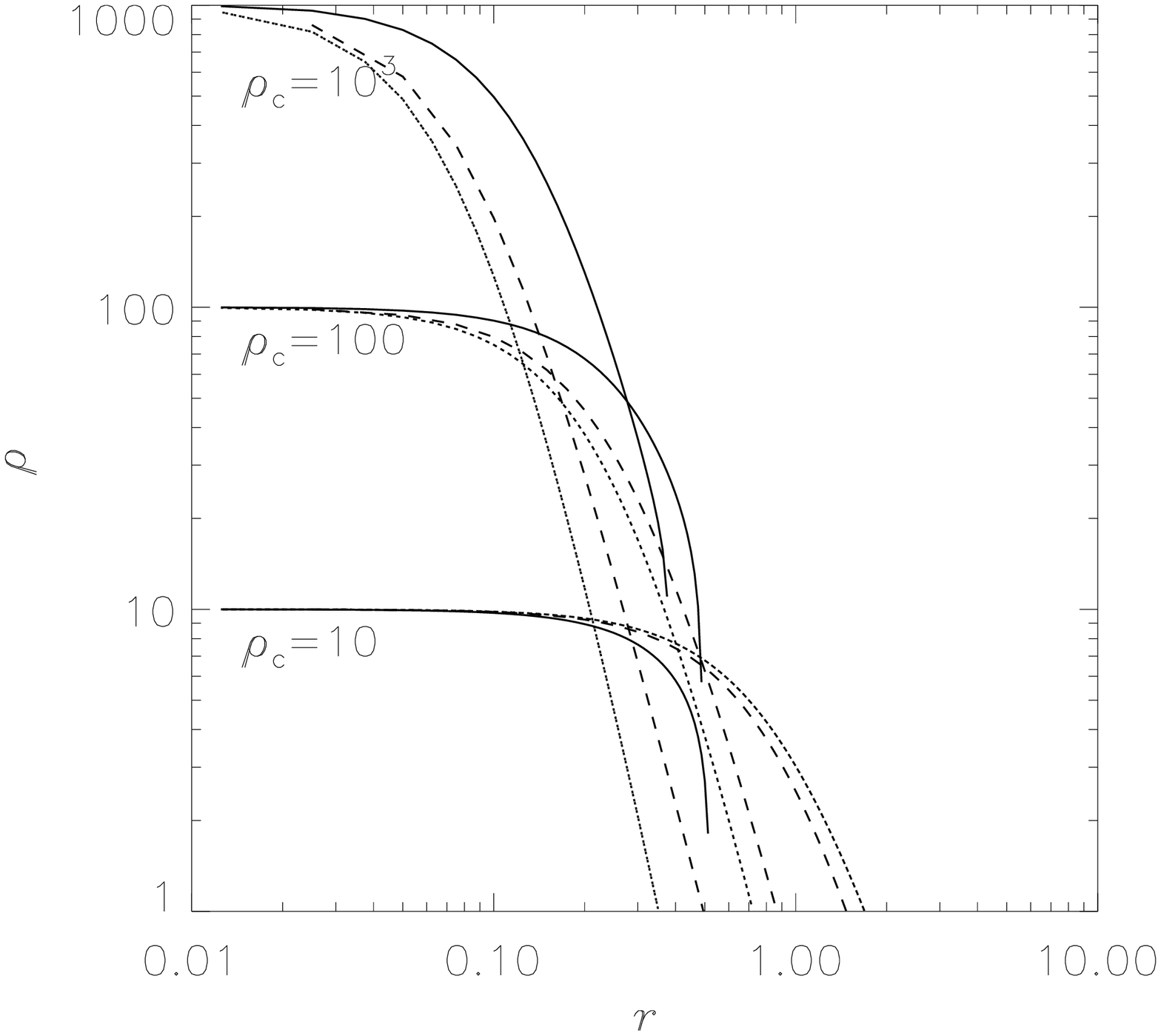}{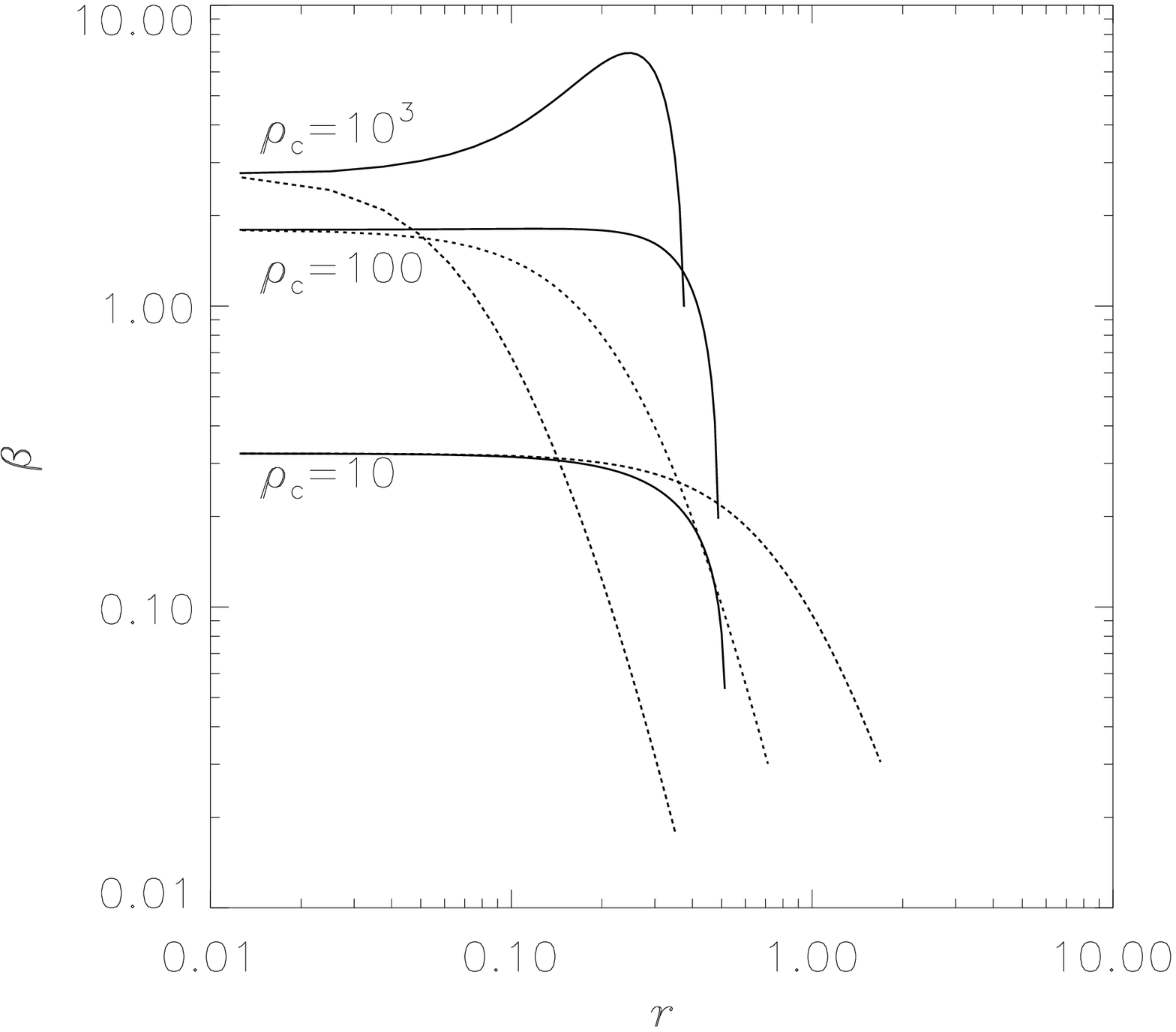}
\caption{Density ({\it a}) and plasma $\beta$ ({\it b}) distributions
 along the $x$- and $y$-axes for Model A.
The horizontal axis represents the distance from the center of the filament,
 $|x|$ or $|y|$. 
Solid and dotted lines represent the distributions respectively along $x$- and $y$-axes.
The dashed lines in ({\it a}) illustrate the density distribution of non-magnetized filament.
\label{fig4}}
\end{figure}

The density and plasma $\beta$ distributions both along the $x$- and $y$-axes
 are shown in Figure \ref{fig4}. 
The distribution in the $y$-direction (dotted line)
 is more extended compared with the $x$-direction (solid line)
 in the model with $\rho_c=10$,
 while the distribution in the $y$-direction (dotted line)
 is more compact than that in the $x$-direction (solid line)
 for $\rho_c=100$ and $10^3$. 
Figure \ref{fig4} ({\it a}) also 
 shows that the density distribution in the $y$-direction
 (dotted line) is similar to that of the non-magnetized filament (dashed line),
 although the distribution is slightly compact compared with the non-magnetized filament.
In addition, the distribution in the $x$-axis is far from that of the non-magnetized filament,
 especially near the surface of the filament.
Although the plasma $\beta$ is small near the surface in this model,
 it increases as it reaches the center.    
Although the plasma beta at the center $\beta_c$ is below unity
 for the models with $\rho_c \lesssim 30$,
 it exceeds unity for the models with $\rho_c \gtrsim 50$.

\subsection{Standard Model}

\begin{figure}
\epsscale{1.0}
\hspace*{2cm}(a)\hspace*{7cm}(b)\\
\plottwo{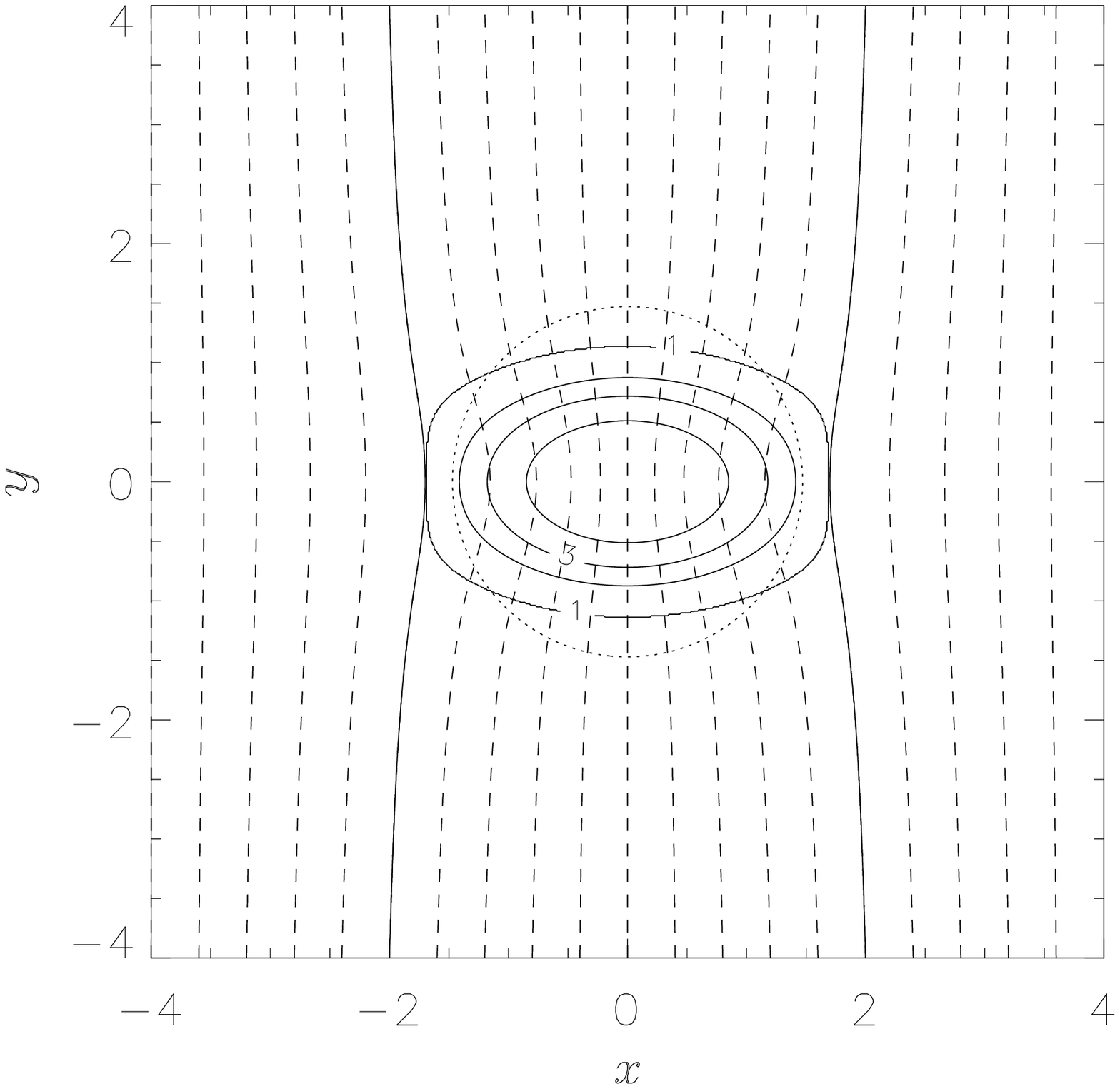}{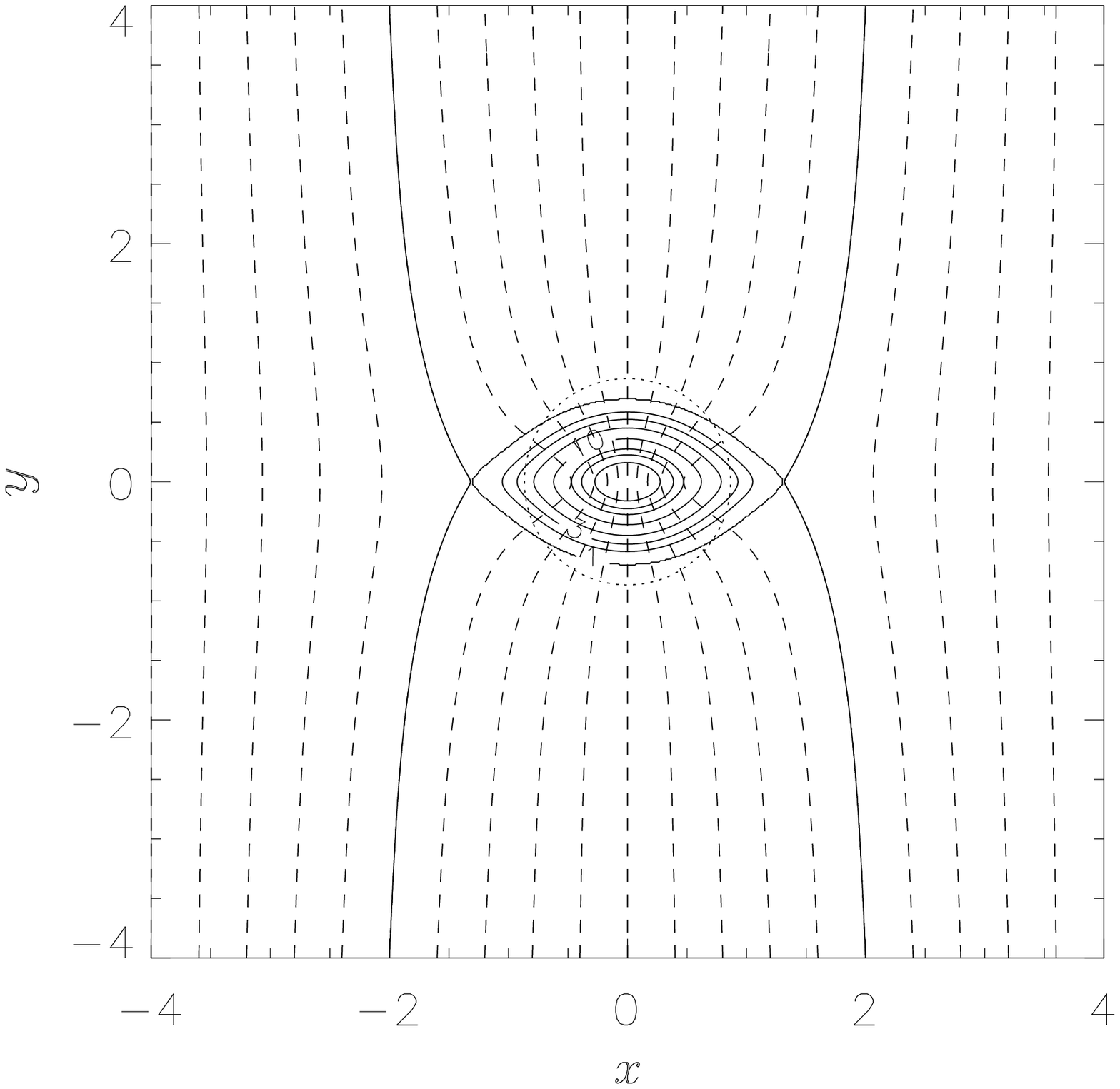}\\
\epsscale{0.5}
\hspace*{4cm}(c)\\
\plotone{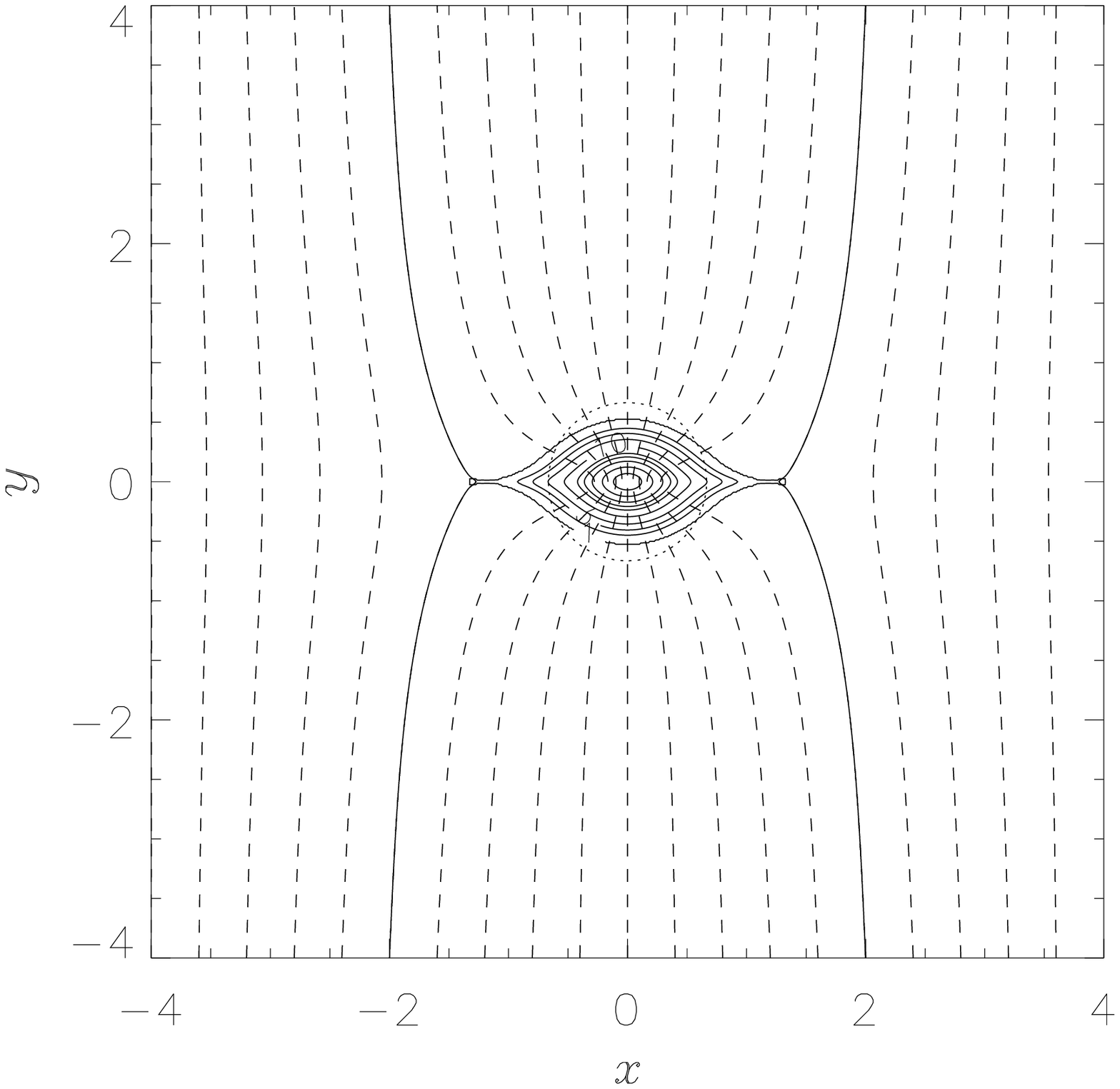}
\caption{Same as Fig.\ref{fig2} but for Model C3 ($R_0=2$ and $\beta_0=1$).
Each panel corresponds to different central densities
 as $\rho_c=10$ ({\it a}), $\rho_c=100$ ({\it b}), and $\rho_c=300$ ({\it c}).
 Each model has a line-mass of $\lambda_0=21.6$ ({\it a}),
 $27.6$ ({\it b}),
 and $28.4$ ({\it c}), respectively.
The model of ({\it c}) has the largest central density as $\rho_c=300$ in Model C3.
\label{fig5}
}
\end{figure}

Now we move on to the models with larger $R_0$ as $R_0=2$ and 5.
In Figure \ref{fig5}, we illustrate the structure of the filament of Model C3
 ($R_0=2$ and $\beta_0=1$) for three central densities, $\rho_c=10$, 100, and 300.
In contrast to Figure \ref{fig2} ({\it a}) and ({\it b}),
 the cross-section of the filament indicates
 a shape whose major axis is perpendicular to the magnetic field.
Panels ({\it b}) and ({\it c}) show that the magnetic field is strongly squeezed 
 inwardly near the mid-plane.
The magnetic field strength is weak in the horizontally peripheral region or in other words
 near the outer mid-plane of the filament.
In the same figure, we plotted the radius of the non-magnetized filament 
 with the same central density by dotted line.
Figure \ref{fig5} indicates that
 the size in the direction parallel to the magnetic field ($y$-direction) is
 slightly smaller than that of the non-magnetized one, while
 that in the perpendicular direction ($x$-direction) is larger
 than that of the non-magnetized one.
In particular, models with high central density (({\it c}): $\rho_c=300$) have 
 a flat shape.

Comparison of models C3 -- C6 shows that
 the half width of the filament in the $y$-direction $Y_s$ 
 decreases with increasing the field strength (from C3 to C6).
That is, models with the same $\rho_c=100$ but different $\beta_0$
 have $Y_s=0.7$ ($\beta_0=1$), 0.65 ($\beta_0=0.5$), 0.55 ($\beta_0=0.1$),
 and 0.475 ($\beta_0=0.01$).
If we compare the respective models for $\rho_c=10$ and for $\rho_c=300$,
 the trend is the same for both central densities. 
On the other hand, that of the $x$-direction $X_s$ increases
 with the magnetic field strength. 
That is, $X_s=1.3$ ($\beta_0=1$), 1.41 ($\beta_0=0.5$), 1.64 ($\beta_0=0.1$),
 and 1.88 ($\beta_0=0.01$), if we compare models with $\rho_c=100$.
This trend is also seen in other central densities. 
This shows that
 the width of the $x$-direction
 \revII{seems to converge} to $X_s \rightarrow R_0=2$ 
 \revII{(the radius of the parent cloud)}
 when increasing the strength of magnetic field.

\begin{figure}
\epsscale{1.0}
\hspace*{2cm}(a)\hspace*{7cm}(b)\\
\plottwo{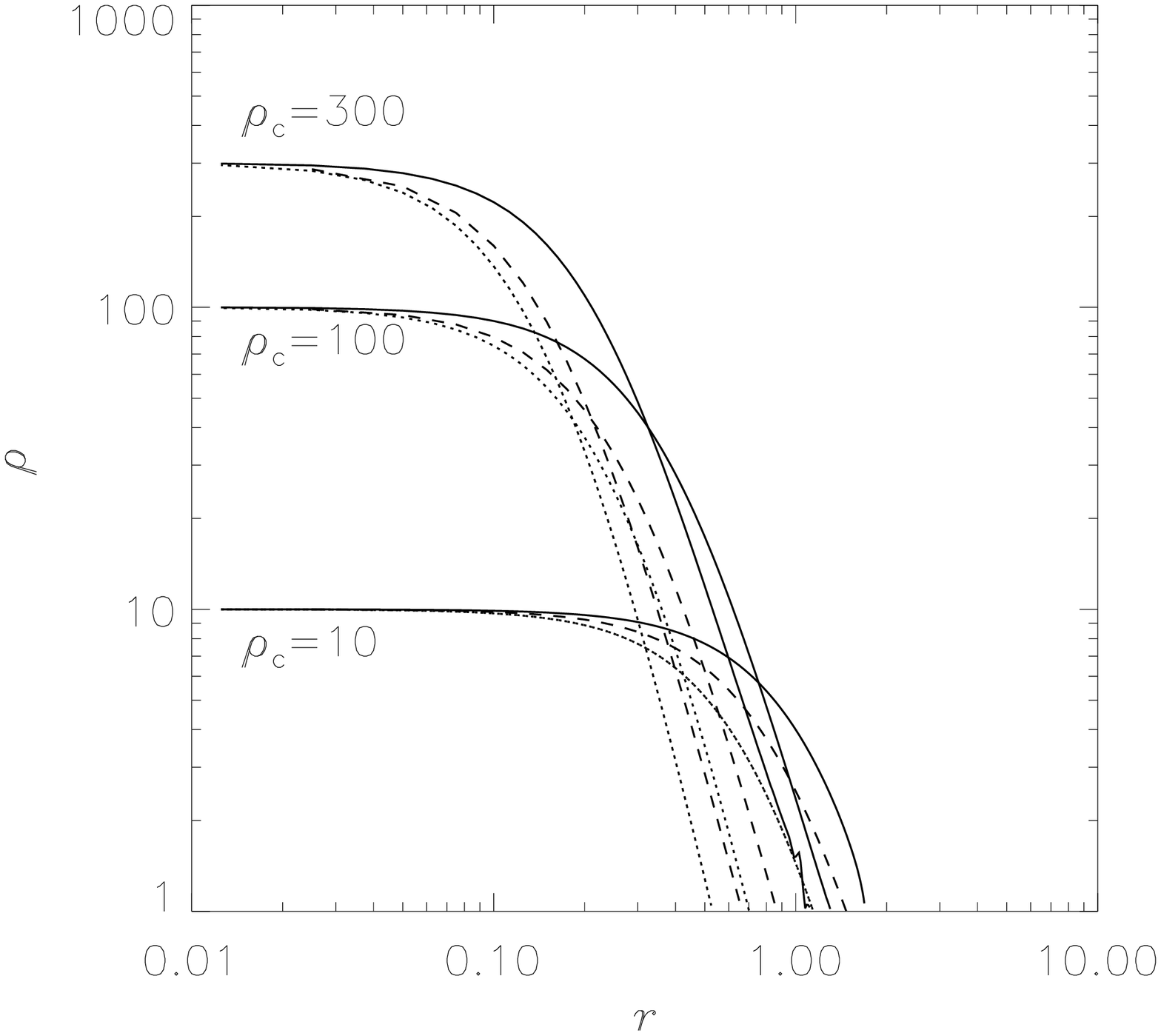}{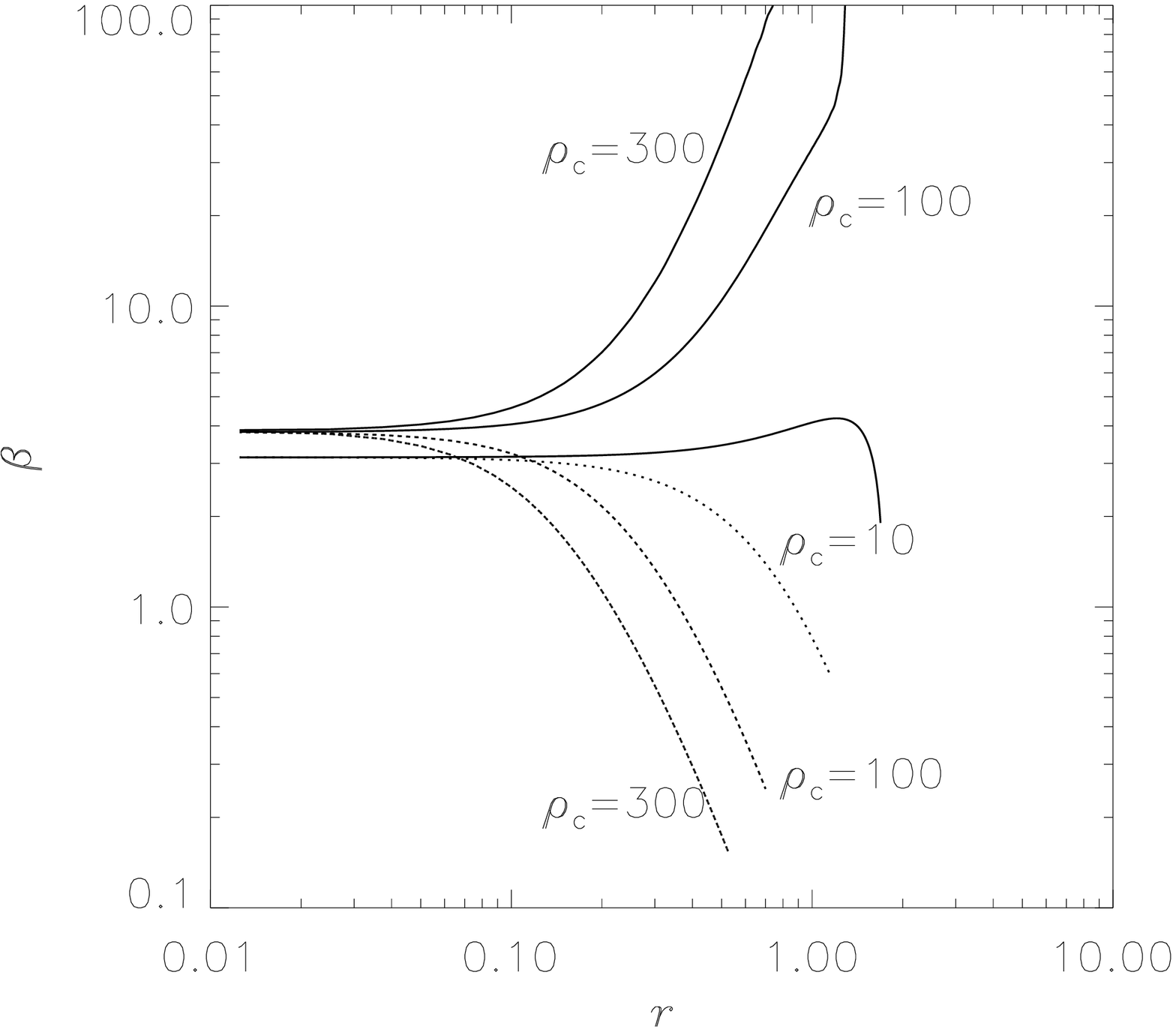}
\caption{Same as Fig.\ref{fig4} but for Model C3 ($R_0=2$ and $\beta_0=1$).
\label{fig6}}
\end{figure}

As shown in Figure \ref{fig6}, the density distributions both in the $x$- (solid line)
 and $y$-directions (dotted line) are more compact for the models with higher central density.
This figure also shows that the slope of the density distribution is 
 similar to that of the non-magnetized filament (dashed line).
The plasma $\beta$ distribution along the $x$- and $y$-axes are illustrated
 in Figure \ref{fig6} ({\it b}).
This figure shows clearly that the central plasma $\beta$ is approximately constant, 
 $\beta_c \simeq 3-4$, irrespective of the central density $10 \le \rho_c \le 300$.
On the other hand, the plasma $\beta$ in the envelope varies greatly depending on
 the central density $\rho_c$.
Furthermore, the plasma $\beta$ increases with increasing distance from the center
 in the $x$-direction, while it decreases in the $y$-direction.
The increase in plasma $\beta$ in the $x$-direction corresponds to the relative weakness
 of the magnetic field seen in the outer ($|x| \gtrsim 0.3-0.5$) mid-plane disk region.
Contrarily, the decrease in plasma $\beta$ in the $y$-direction is explained 
 by the decrease in the density (and simultaneously the pressure). 
We should pay attention to the great contrast to the models with small $R_0$ 
 of Figure \ref{fig4} ({\it b}).
The plasma $\beta$ decreases near the surface in both $x$- and $y$-directions for Model A.
The inhomogeneity in the magnetic field is shown to be characteristic of the models
 with larger $R_0$.

\begin{figure}
\epsscale{1.0}
\hspace*{2cm}(a)\hspace*{7cm}(b)\\
\plottwo{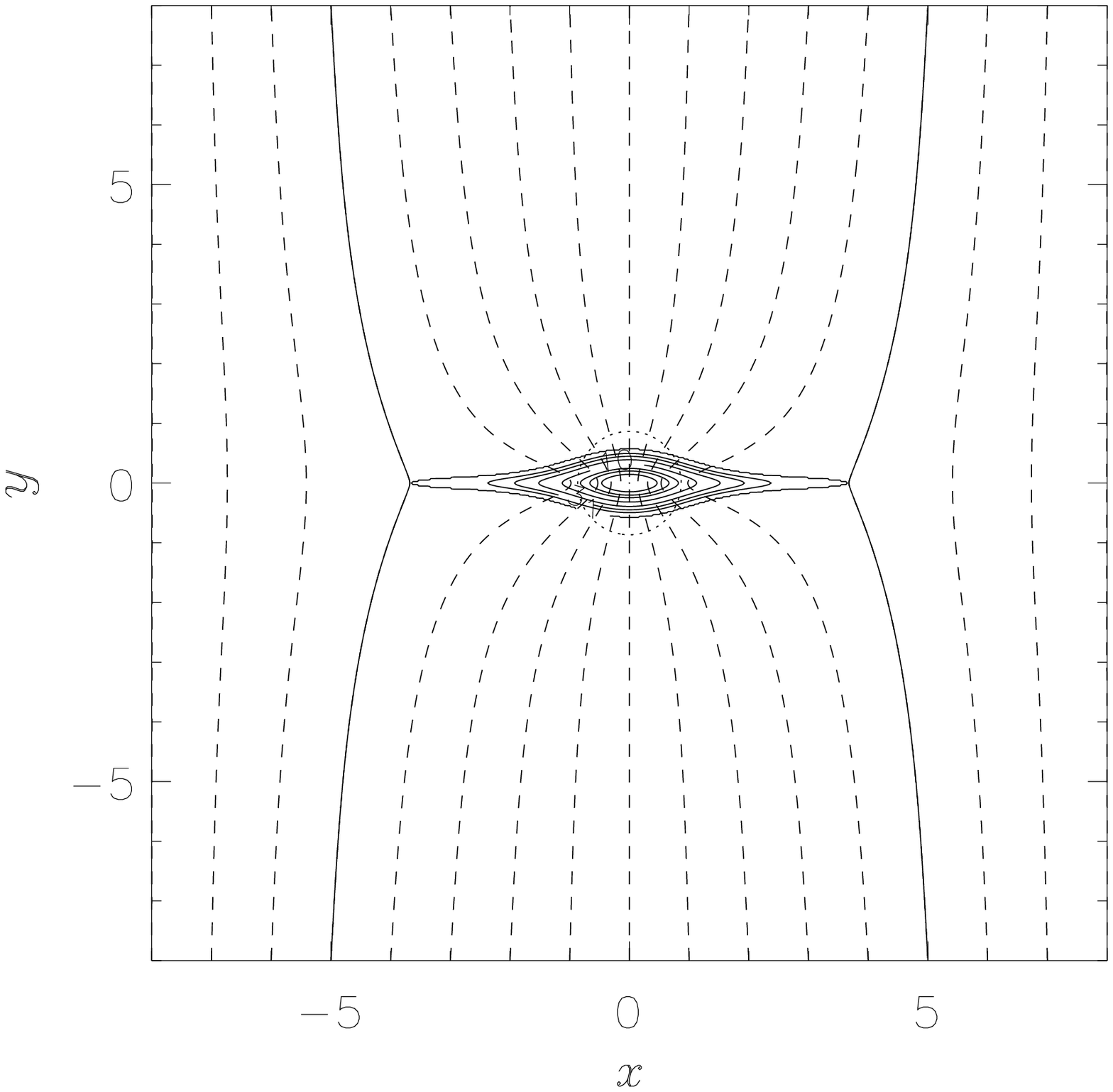}{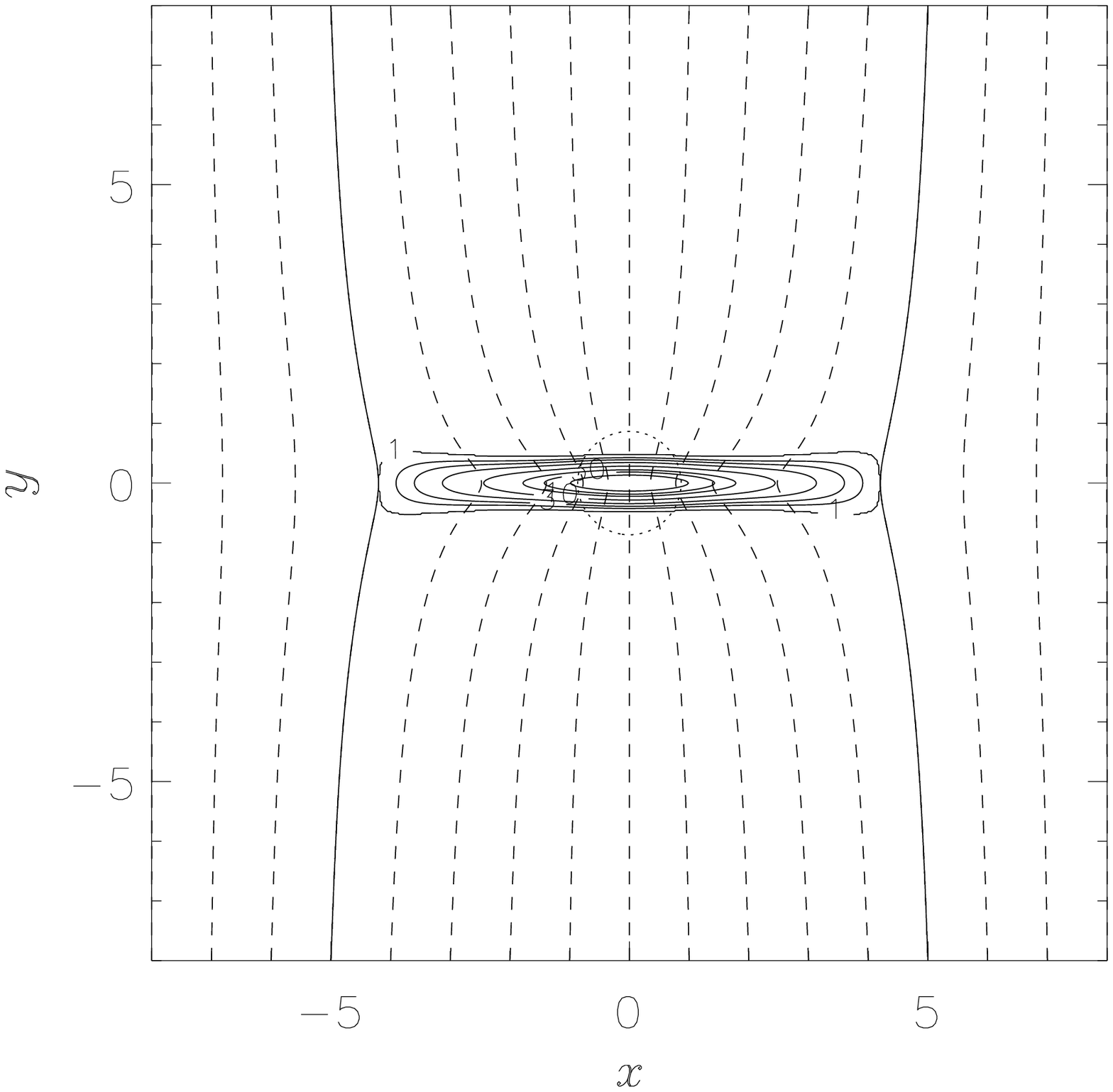}\\
\epsscale{0.5}
\hspace*{4cm}(c)\\
\plotone{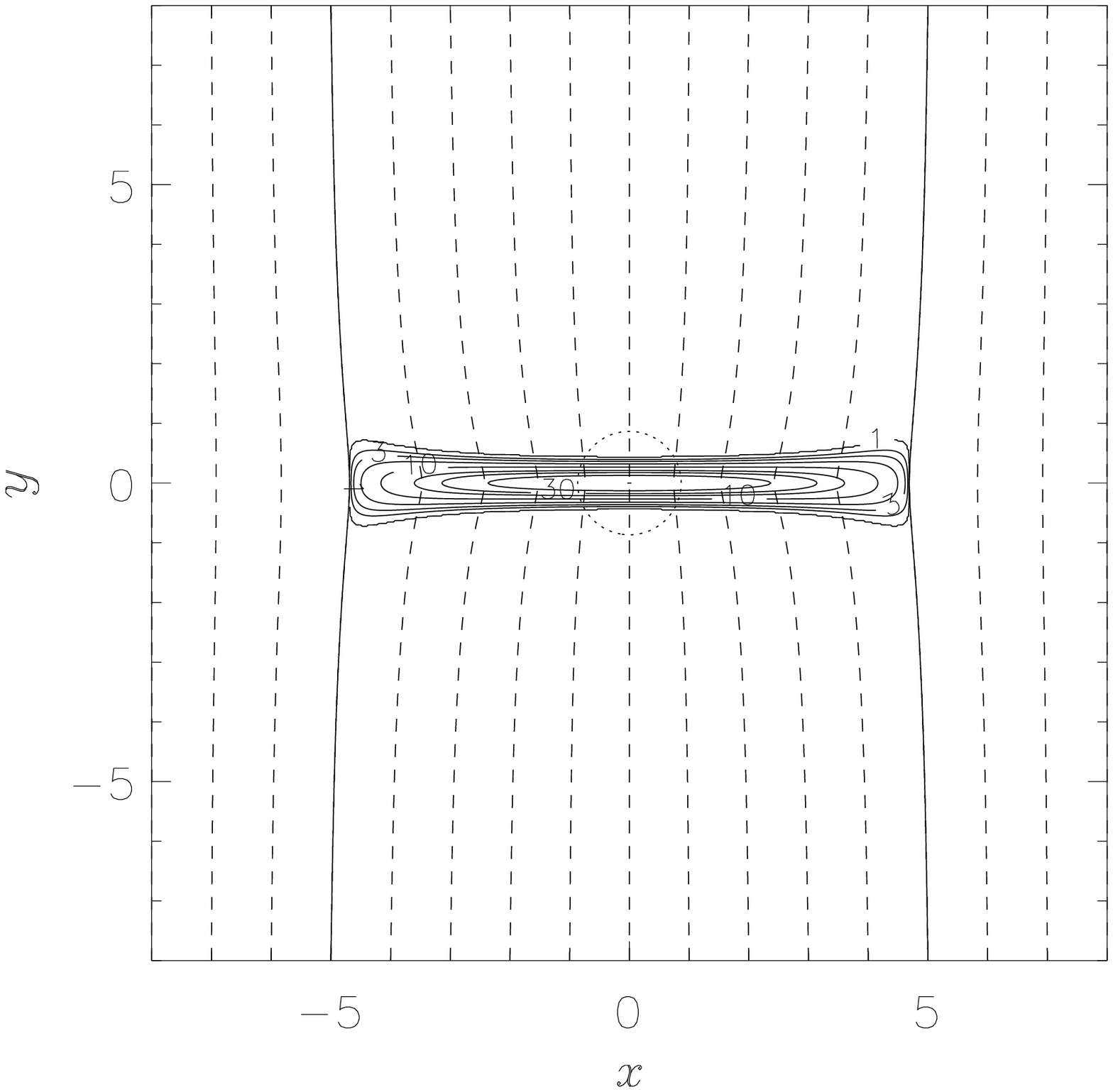}
\caption{Same as Fig.\ref{fig2} but for Models D ($R_0=5$).
Each panel corresponds to different $\beta_0$ but the same $\rho_c=100$:
 Models D1 ({\it a}), D2 ({\it b}), and D3 ({\it c}) correspond respectively
 to $\beta_0=1$, $\beta_0=0.1$, and $\beta_0=0.01$.
 Each model has a line-mass of $\lambda_0=40.8$ ({\it a}),
 $85.9$ ({\it b}),
 and $169$ ({\it c}), respectively.
\label{fig7}
}
\end{figure}

\begin{figure}
\epsscale{1.0}
\hspace*{2cm}(a)\hspace*{7cm}(b)\\
\plottwo{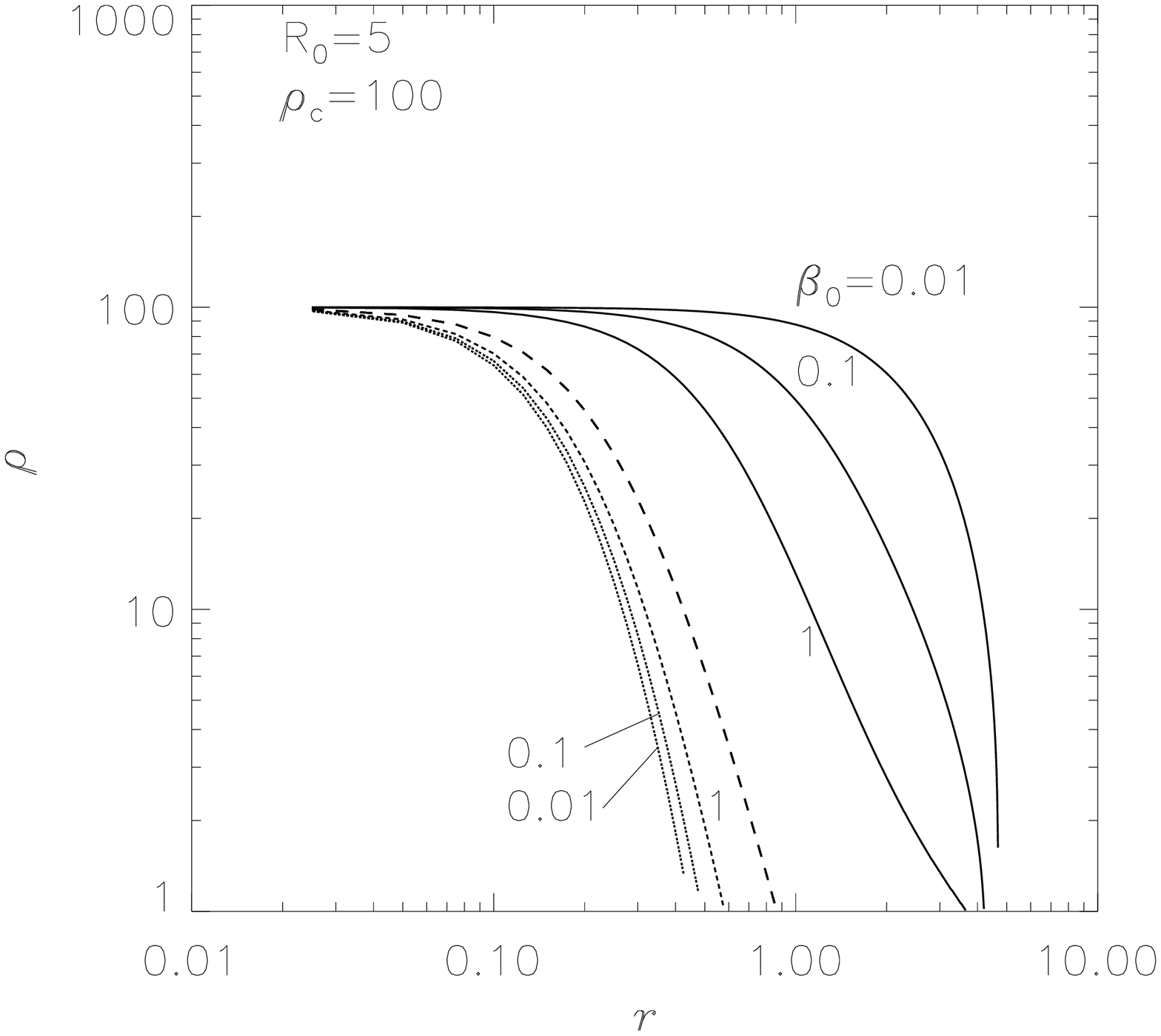}{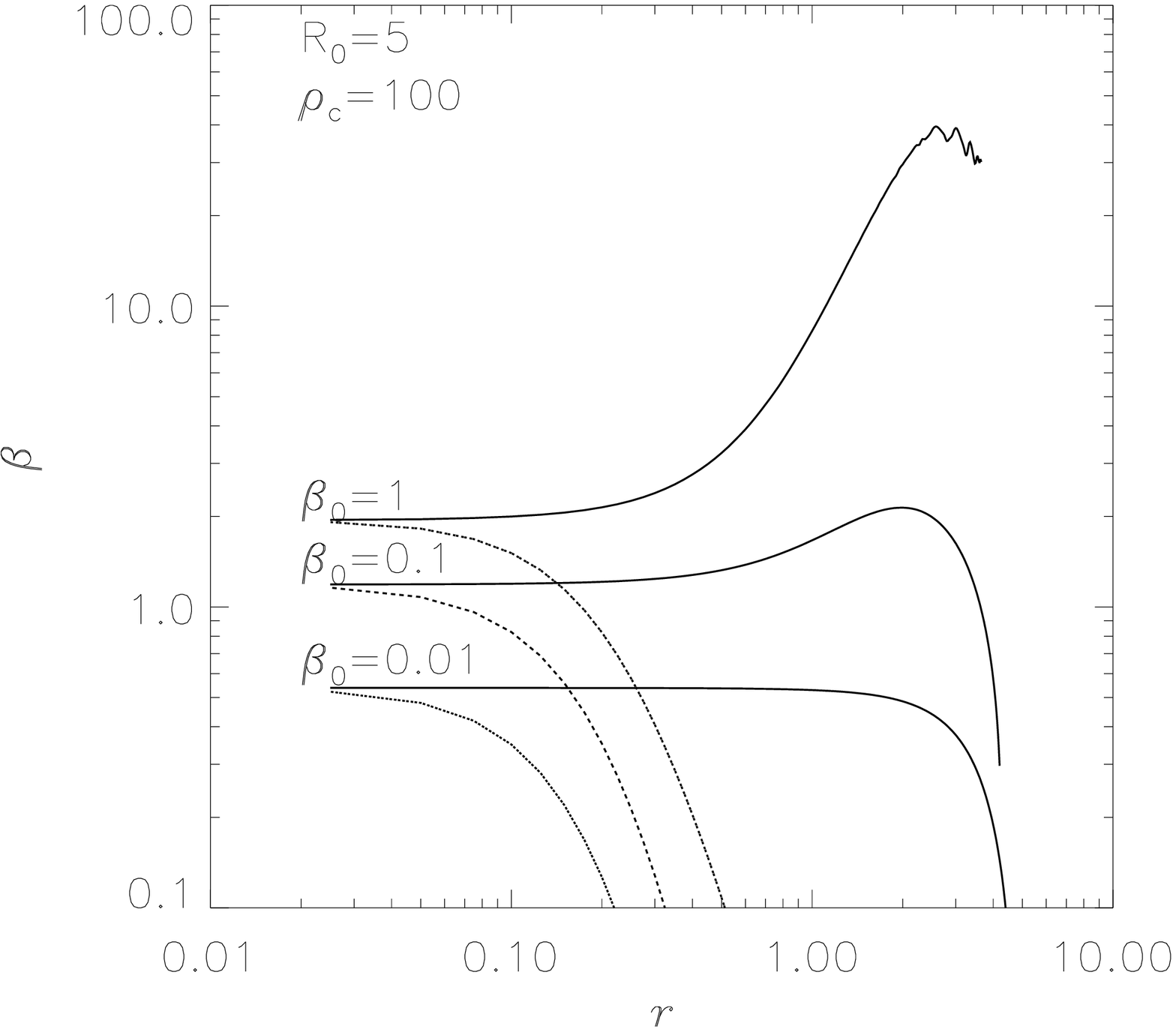}
\caption{Same as Fig.\ref{fig4} but for Models D1, D2, and D3 ($R_0=5$).
Comparison is made between the models with the same central density $\rho_c=100$
 but different $\beta_0=1$, 0.1, and 0.01.
\label{fig8}}
\end{figure}

Figure \ref{fig7} illustrates the structure of Models D1 ({\it a}), D2 ({\it b}), and D3 ({\it c}).
Comparing Models C3 (Fig. \ref{fig5}({\it b})) and D1 (Fig. \ref{fig7}({\it a})) 
 under the condition of $\rho_c$ being fixed,
 we can see that the major-to-minor axis ratio of the cross-section of
 filaments increases with $R_0$.
This seems to come from the fact that
 the size in the $x$-direction is approximately given by the initial radius $R_0$
 but the width of the $y$-direction is similar to the diameter of the non-magnetized
 filament.
Increasing the magnetic field strength from $\beta_0=1$ ({\it a}) to
 $\beta_0=0.01$ ({\it c}), curvature of the magnetic field decreases,
 since the strong magnetic field prevents itself from  bending.
Although the horizontal size of the filament is similar,
 these three models have completely different line-masses 
 $\lambda_0=40.8$ ({\it a}),
 $85.9$ ({\it b}),
 and $169$ ({\it c}), respectively.
Figure \ref{fig8}({\it a}) plots the density distributions along the $x$- and $y$-axes.
The distribution along the $x$-axis (the solid line) is extended in the model with stronger
 magnetic field (Model D3 $\beta_0=0.01$). 
On the other hand, Model D1 indicates a centrally condensed density distribution 
 and the slope of $\rho(|x|)$ is shallower than that of the non-magnetized filament
 (the dashed line) in the envelope region $r\gtrsim 1$.   
The figure also indicates that the distribution in the $y$-direction is more compact than
 that of the non-magnetized filament with the same central density, which is also seen  
 in Models C.

As seen in Figure \ref{fig8} ({\it b}), 
the central plasma $\beta$ varies as $\beta_c \simeq  2.1$ 
 ({\it a}: $\beta_0=1$),
 1.2 ({\it b}: $\beta_0=0.1$), and 0.54 ({\it c}: $\beta_0=0.01$),
 according to $\beta_0$.
That is, increasing $\beta_0$ in two orders of magnitude induces the increase
 of a factor of $\sim 4$ in $\beta_c$.
Although the plasma $\beta$ decreases along the $y$-axis
 with increasing distance from the center (dotted line),
 distribution of the plasma $\beta$ along the $x$-axis is complex (solid line).
In Model D1 ({\it a}), $\beta(|x|)$ increases from $\sim 2$ to $40$ along the $x$-axis;
 in Model D2 ({\it b}), $\beta(|x|)$ increases from $\simeq 1.2$ and reaches $\simeq 2$ then
 reduces below unity;
 in Model D3 ({\it c}), $\beta(|x|)$ decreases from $\simeq 0.5$ monotonically.
This explains the fact that the magnetic field is extremely weak in the outer region of 
 the filament, $|x| \gtrsim 3$ in Model D1, while the field strength is almost uniform
 in Model D3 where the Lorentz force is much stronger than the pressure force
 and gravity.

We illustrate the relation between the line-mass and the central
 density for Models C and D ($R_0=2$ and 5) in Figure \ref{fig3}({\it b}),
 which shows that the line-mass of the magnetized filaments
 is always larger than that of the non-magnetized one (dash-dotted line). 
This means that the magnetic field plays a role in supporting the isothermal
 filament against the self-gravity.
In contrast to Models C and D, Models A and B with small $R_0$ show that 
 the magnetic field has an effect to confine the filamentary gas
 and thus the line-mass of magnetized  filament is sometimes less-massive
 than that of the non-magnetized filament.
 
As a result, the line-masses of Models C and D (standard model)
 are more massive than those of Models A and B (the models with small $R_0$),
 comparing between the respective models with the same central density.

\section{Discussion}

\subsection{Maximum Mass}
\label{sec:4.1}
\begin{figure}
\plotone{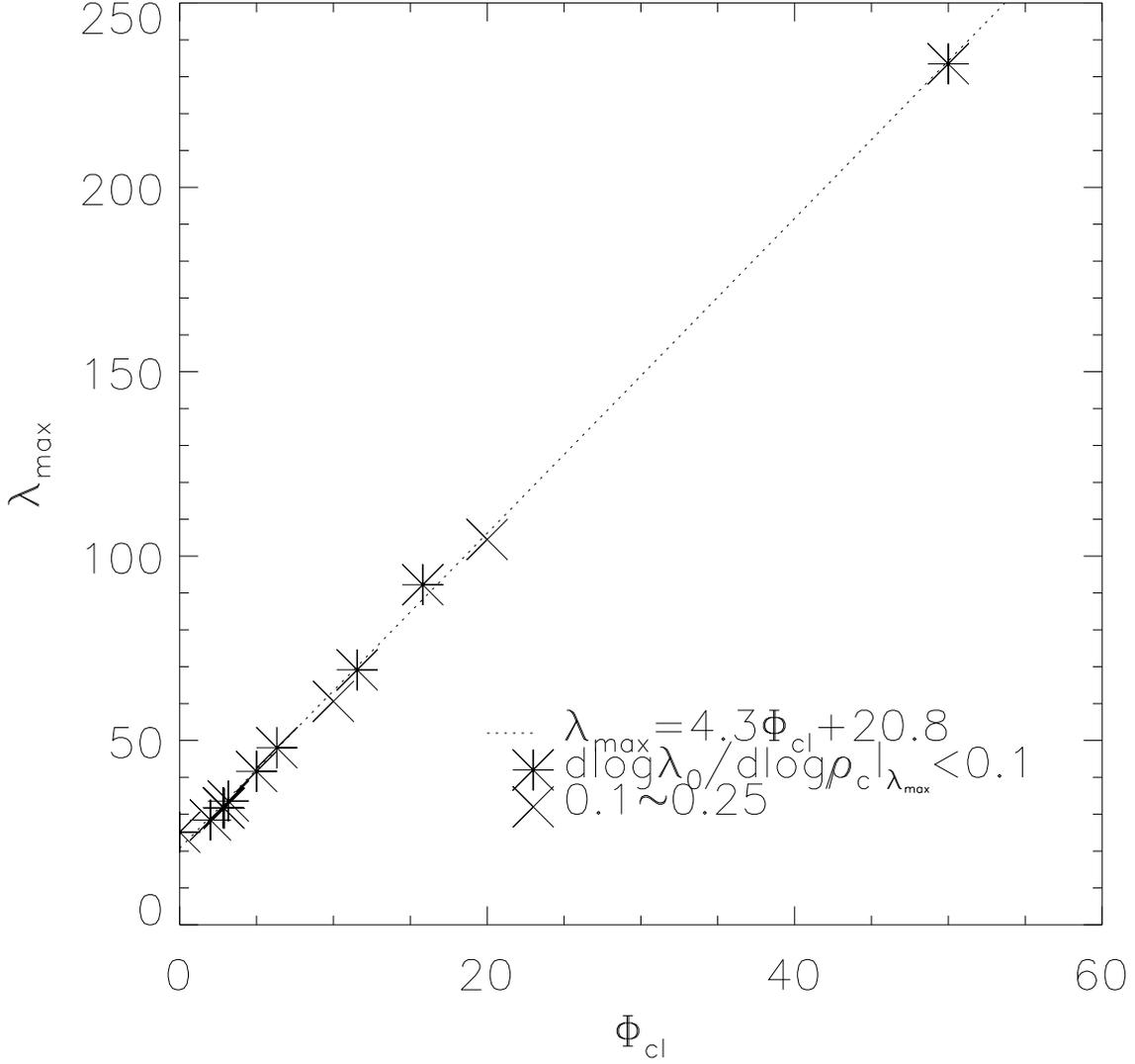}
\caption{Maximum line-mass is plotted against the flux.
Since the supported line-mass seems
 an increasing function of the central density, 
 the maximum line-mass obtained for $\rho_c\lesssim 10^3$
 is actually a lower limit.
\revII{Symbols denote} the value of $\dif{\log \lambda_0}{\log\rho_c}$
 at the maxima:
 the asterisk represents $\dif{\log \lambda_0}{\log\rho_c}\le 0.1$,
 which gives more exact critical line-mass than the others, while
 the cross represents $0.1<\dif{\log \lambda_0}{\log\rho_c}\le 0.25$.
From the asterisk points the critical line-mass is fitted
 as $\lambda_{\rm max}=4.3\Phi_{\rm cl}+20.8$ (dotted straight line). 
\label{fig9}}
\end{figure}

Figure \ref{fig3} shows the masses of the equilibrium solutions.
As is expected from the non-magnetized model,
 the line-mass is an increasing function of the central density or central-to-surface
 density ratio.
The non-magnetized filament has a maximum line-mass of
 $\lambda'_0=8\pi$ (see eq.[\ref{eqn:lambda-nondim}]),
 which corresponds to the dimensional value of $\lambda_0=2c_s^2/G$
 (see eq.[\ref{eqn:lambda-dim}]), which is reached for
 $\rho_c\rightarrow \infty$.
Similarly to the non-magnetized model,
 the maximum line-mass for given parameters $R_0$ and $\beta_0$ 
 seems to be reached when $\rho_c\rightarrow \infty$. 
The maximum line-mass should be estimated from the line-mass of the filament with finite central densities.
We calculate $\dif{\log \lambda_0}{\log \rho_c}$ for the state with the highest
 central density.
When the slope $\dif{\log \lambda_0}{\log \rho_c}$ is as small as $ < 0.1$,
 the maximum line-mass given from the largest line-mass obtained is a good 
 estimation as the maximum line-mass that can be supported.
The largest line-masses obtained for respective models are plotted in Figure \ref{fig9} and shown in Table \ref{table1}.
The line-mass ($y$-axis) is plotted against the magnetic flux per unit length $\Phi_{\rm cl} \equiv R_0B_0$ ($x$-axis).   
Asterisks represent the maximum line-mass for more reliable models,
 as   $\dif{\log \lambda_0}{\log \rho_c} \le 0.1$, while the cross represents
 the maximum line-mass for the models with $0.1 < \dif{\log \lambda_0}{\log \rho_c} \le 0.25$.
From the asterisk points, we obtain an empirical formula for the maximum line-mass
 as
\be
\lambda'_{\rm max} \simeq 4.3 \Phi_{\rm cl}' + 20.8,
\label{eqn:-lambda'}
\ee  
using the least-squares method, where we add $'$ to emphasize that the quantities are
 normalized.
Although the maximum line-mass of the non-magnetized filament given from the formula  
 $\lambda'_{\rm max}(\Phi_{\rm cl}'=0)\simeq 20.8$ is somewhat smaller than that obtained 
 analytically $\lambda'_{\rm max}(\Phi_{\rm cl}'=0)=8\pi\simeq 25.1$,
 the fit is remarkable. 
Since the line-mass and the magnetic line-flux are normalized by
 $c_s^2/4\pi G$ and $c_s^2/(G/2)^{1/2}$ in this paper, the dimensional maximum line-mass
 is expressed with the dimensional flux as
\be
\lambda_{\rm max} \simeq 0.24 \frac{\Phi_{\rm cl}}{G^{1/2}}+1.66 \frac{c_s^2}{G}.
\label{eqn:mcr_nond}
\ee 
The critical mass-to-flux ratio has been obtained for the disk-like cloud as
 $(G^{1/2}M/\Phi_B)_{\rm crit}\simeq 0.18 \simeq 1/2\pi$ (eq.(4.1) of \citet{tomisaka1988b}; see also \citet{nakano78}).
It should be noticed that the factor for the filament, $0.24$,
 is similar to that of the disk, $0.18$\footnote{
The mass-to-flux ratio ($M/\phi$) and the line-mass to flux per unit 
length ($\lambda/\Phi_{\rm cl}$) ratio have the identical dimension to the ratio
 between column density and magnetic flux density ($\sigma/B$).  
Increasing the strength of magnetic field, a flat structure appears
 near the center of the cloud both in a disklike cloud \citep{tomisaka1988b} and in a filamentary cloud.  
Thus, a flat structure threaded perpendicularly by the magnetic 
 field appears commonly
 and such a disk seems to control the maximum mass and line-mass.  
If the condition of criticality is related to the structure of this flat 
 part and thus related to the $\sigma/B$ ratio of the structure,
 this explains the similarity of the factors appeared  
 in the critical $M/\phi$ and in the critical $\lambda/\Phi_{\rm cl}$.}.
Finally, the maximum line-mass (eq.[\ref{eqn:mcr_nond}]) is finally evaluated as
\be
\lambda_{\rm max} \simeq 22.4  \left(\frac{R_0}{0.5\,{\rm pc}}\right)
                             \left(\frac{B_0}{10\,{\rm \mu\, G}}\right) M_\odot\,{\rm pc}^{-1}
+13.9\left(\frac{c_s}{190\, {\rm m\,s^{-1}}}\right)^2 M_\odot\,{\rm pc}^{-1}.
\label{eqn:mcr_dim}
\ee 
Thus, when the magnetic flux per unit length is larger than
 $\Phi \gtrsim 3\, {\rm pc \,\mu G}(c_s/190\,{\rm m\,s^{-1}})^2$,
 it is concluded
 that the maximum line-mass of the filament is affected by the magnetic field
 (the first term is larger than the second term).

The factors in front of $\Phi'_{\rm cl}$ and $\Phi_{\rm cl}$ in equations 
 (\ref{eqn:-lambda'}) and (\ref{eqn:mcr_nond})
 must depend on the way of mass loading adopted here
 (eq.[\ref{eqn:model_dm_dPhi}]).
Analogy from the case of disk-like cloud (\S IVb of \citet{tomisaka1988b}),   
 the factor may increase if we choose more uniform $d\lambda/d\Phi_{\rm cl}$ rather than the centrally concentrated one assumed in equation (\ref{eqn:model_dm_dPhi}).

\subsection{Virial Analysis}
In this section,
 we analyze the equilibrium state of the cloud using a virial analysis.
The equation of motion for MHD is written
\be
\rho\left[\difd{\bf u}{t}
         +\left({\bf u}\cdot\nabla\right){\bf u}\right]=-\nabla p -\rho\nabla\psi 
 +\frac{1}{4\pi}\left(\nabla\times {\bf B}\right)\times {\bf B},\label{eqn:eq-motion}
\ee
where {\bf u} represents the flow velocity.
We consider an axisymmetric filament uniform in the axis direction.
\revII{Using the cylindrical coordinates,
 we multiply the position vector from the center,
 and integrate over the filament.
The first term of the right-hand side gives}
\be
\int_0^R -\difd{p}{r}r2\pi r dr=-\left[p 2 \pi r^2\right]_0^{R}+\int_0^Rp4\pi rdr
=-2\pi R^2p_s+2c_s^2\lambda,\label{eqn:virial1}
\ee 
where $R$, $p_s$, and $\lambda$ represent the radius of the filament,
 the pressure at the surface of the filament, and the line-mass
 defined as $\lambda=\int_0^R \rho 2\pi r dr$.
Treating the second term of the right-hand side of equation (\ref{eqn:eq-motion})
 in a similar way provides 
\be
\int_0^R -\rho \difd{\psi}{r}r 2 \pi rdr
=-\int_0^R 2G\lambda_r \rho 2\pi r dr
=-2G\int_0^R \lambda_r d\lambda_r
=-G\lambda^2, \label{eqn:virial2}
\ee
where we use the Poisson equation for the gravity as
$
-\dif{\psi}{r}=-{2G\lambda_r}/{r}
$
and the definition of the line-mass contained inside the radius $r$ as
$
\lambda_r=\int_0^r \rho 2\pi r dr.
$
Thus, for a non-magnetic filament ${\bf B}=0$,
equation (\ref{eqn:eq-motion}) gives the virial equation in equilibrium
\be
2\pi R^2 p_s = G \lambda\left(\frac{2c_s^2}{G}-\lambda\right)
\label{eqn:eq-of-B=0}
\ee
Since the right-hand side of the equation must be positive,
 the line-mass $\lambda$ has a maximum, $\lambda_{\rm max}=2c_s^2/G$.
This critical line-mass is estimated as 
 $\lambda_{\rm max}\simeq 17\, M_\odot{\rm pc^{-1}}(c_s/190\,{\rm m\,s^{-1}})^2$
 for the interstellar molecular gas with $T=10\, {\rm K}$.

When the filament is laterally threaded by the magnetic field,
 the last term of the right-hand side of equation (\ref{eqn:eq-motion})
 gives the magnetic force term and is estimated as
\be
\frac{B^2}{8\pi}\pi R^2 =\frac{\Phi_{\rm cl}^2}{8},
\label{eqn:magnetic-term}
\ee 
where $\Phi_{\rm cl}$ represents the magnetic flux per unit length as
$\Phi_{\rm cl}=BR=B_0R_0$.
Here we assumed that the flux $\Phi_{\rm cl}$ is \revII{the same as in the 
 parent state} in which the radius and the magnetic field strength 
 are equal to $R_0$ and $B_0$, respectively.
Equation (\ref{eqn:eq-of-B=0}) becomes 
\be
2\pi R^2 p_s = 2c_s^2\lambda-G\lambda^2+\frac{\Phi_{\rm cl}^2}{8}.
\label{eqn:eq-of-B<>0}
\ee
Thus, the maximum line-mass increases and is equal to
\be
\lambda=\frac{c_s^2+(c_s^4+G\Phi_{\rm cl}^2/8)^{1/2}}{G}
\sim \frac{2c_s^2}{G}+\frac{\Phi_{\rm cl}^2}{16c_s^2},
\label{eqn:mass-formula-small_phi}
\ee
where to derive an approximate formula \revII{for} $\lambda$ \revII{in the weakly magnetized
 case} we assumed $2c_s^2/G\gg \Phi_{\rm cl}^2/(16c_s^2)$ or 
 $\Phi_{\rm cl} \ll 3\, {\rm pc\,\mu G}(c_s/190\, {\rm m\,s^{-1}})^{2}$.
The extra line-mass allowed by the magnetic field is expected to be 
\be
\Delta \lambda \sim 2.5\, M_\odot\,{\rm pc}^{-1} (R_0/0.1\,{\rm pc})^2(B_0/10\,{\rm \mu G})^2(c_s/190\,{\rm m\,s^{-1}})^{-2}.
\ee
Equation (\ref{eqn:mass-formula-small_phi})
 means that the maximum line-mass is expressed as
 $\lambda'_{\rm max}=a\Phi_{\rm cl}'^2 + b$ in the limit of small magnetic flux per unit length.
Using our result for the five models with $0 \le \Phi_{\rm cl}' =R'_0/\beta_0^{1/2} < 5$
 (Models A, B1, C3, C4 and the non-magnetized model),
  the maximum line-mass is well fitted by the following formula as
\be
 \lambda'_{\rm max}=0.82 \Phi_{\rm cl}'^2 +25,
\ee
or in the dimensional form as
\be
 \lambda_{\rm max}=0.033 \Phi_{\rm cl}^2/c_s^2 + 2.0 c_s^2/G.
\label{eqn:empirical-lambda-weak-B}
\ee
Although the numerical factor in front of $\Phi_{\rm cl}^2$ is twice
 as large as that expected from the virial analysis,
 our empirical fit [eq.(\ref{eqn:empirical-lambda-weak-B})]
 seems to have a theoretical base.

In the case of $2c_s^2/G \ll \Phi_{\rm cl}^2/(16c_s^2)$ or $\Phi_{\rm cl} \gg 3\, {\rm pc\,\mu G}(c_s/190\, {\rm m\,s^{-1}})^{2}$, equation (\ref{eqn:eq-of-B<>0}) gives
\be
\lambda \simeq \frac{c_s^2}{G}+\frac{\Phi_{\rm cl}}{2^{3/2}G^{1/2}}+\frac{2^{1/2}c_s^4}{G^{3/2}\Phi_{\rm cl}},
\ee
and the mass increment due to the magnetic field (the second term)
 is expected to be proportional to
 the magnetic flux per unit length
 for a filament with large $\Phi_{\rm cl}$ as follows:
\be
\Delta \lambda \sim 30\, M_\odot\,{\rm pc}^{-1} (R_0/0.5\, {\rm pc})(B_0/10\, {\rm \mu G}).
\ee
This  reproduces the empirical mass formula obtained numerically (eq.~[\ref{eqn:mcr_dim}]).
When the filament is magnetically supported,
 the maximum line-mass is proportional to the magnetic flux per unit length $\Phi_{\rm cl}$.
Therefore, the functional form of the empirical mass formula in $\S$\ref{sec:4.1} also has a theoretical meaning.

\subsection{Column Density Distribution}

As for the modeling of the density profile of the filament,
Plummer-like profiles are commonly adopted as \citep{nutter2008,arzoumanian2011}
\be
\rho(r)=\frac{\rho_c}{[1+(r/R_f)^2]^{p/2}},
\ee
which has the analytic expression also for the column density as
\be
\sigma(r)=A\frac{\rho_cR_f}{[1+(r/R_f)^2]^{(p-1)/2}},
\label{eqn:plummer-sigma}
\ee
where $A$ is a numerical factor.
The distribution is determined by the central density $\rho_c$,
 core radius $R_f$, and the density slope parameter $p$ 
 \citep{nutter2008,arzoumanian2011}.
However, there is a contradiction between theory and observation.
That is, Herschel observations indicate $p \sim 2$, although the 
 isothermal hydrostatic filament has $p=4$ (eq.[\ref{eqn:cyl-rho}]).
This is sometimes considered as a consequence of dynamical contraction.    
Dynamical contraction of a filament whose pressure obeys the polytropic 
relation $p\propto \rho^\gamma$ is expressed by a self-similar solution
and has a slope such as $\rho \propto r^{-2/(2-\gamma)}$ \citep{kawachi1998}.
And the infall speed expected from the solutions is reported consistent 
with observations \citep{palmeirim2013}.
On the other hand, from a standpoint that the filament is hydrostatic,
 some ideas to change the gas equation-of-state are proposed.
Here we study the possibility that the column density distribution is
 changed by a magnetic effect.
We study the column density distribution expected for the
 magnetized filament.

\begin{figure}
\epsscale{1.0}
\plottwo{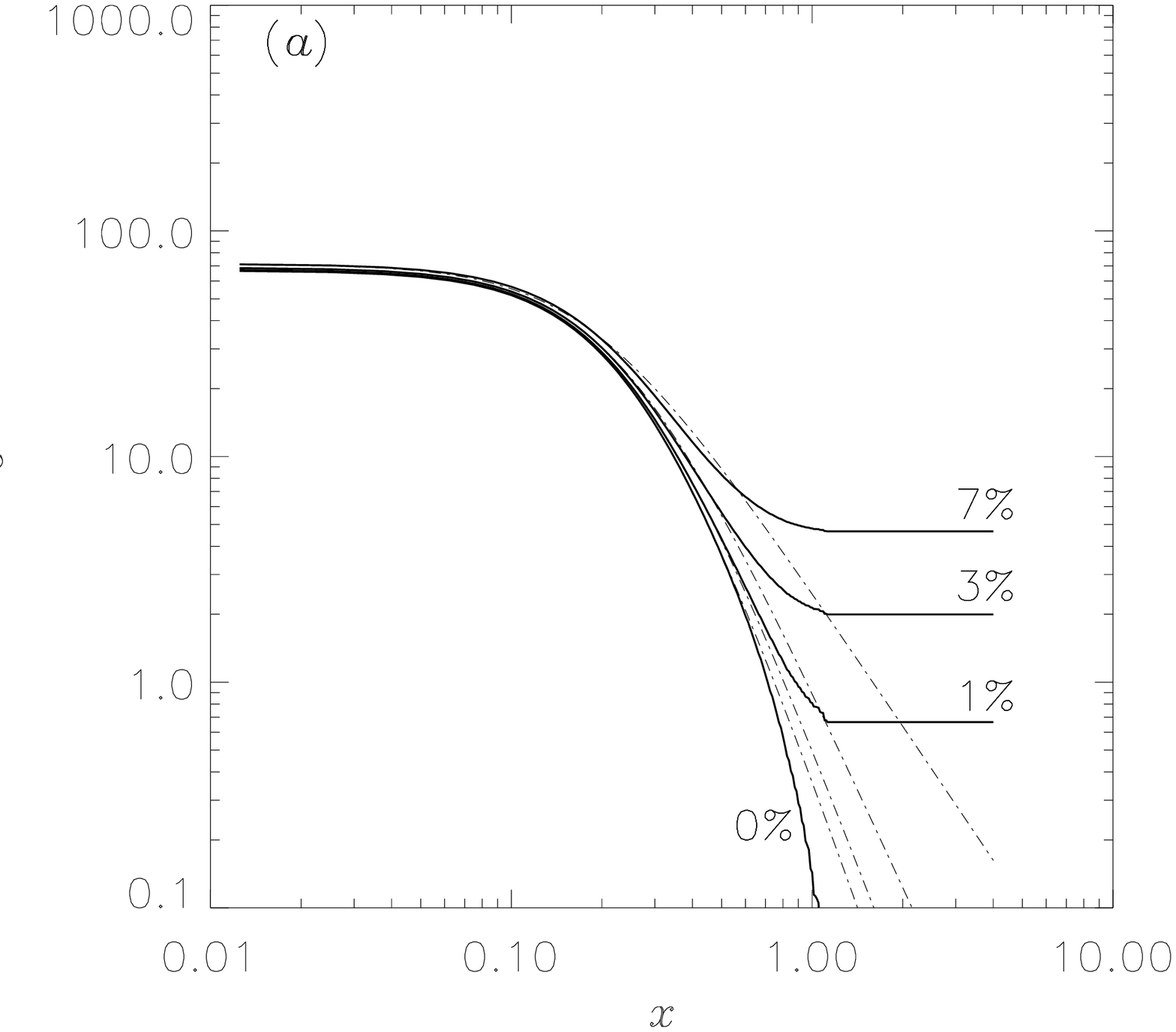}{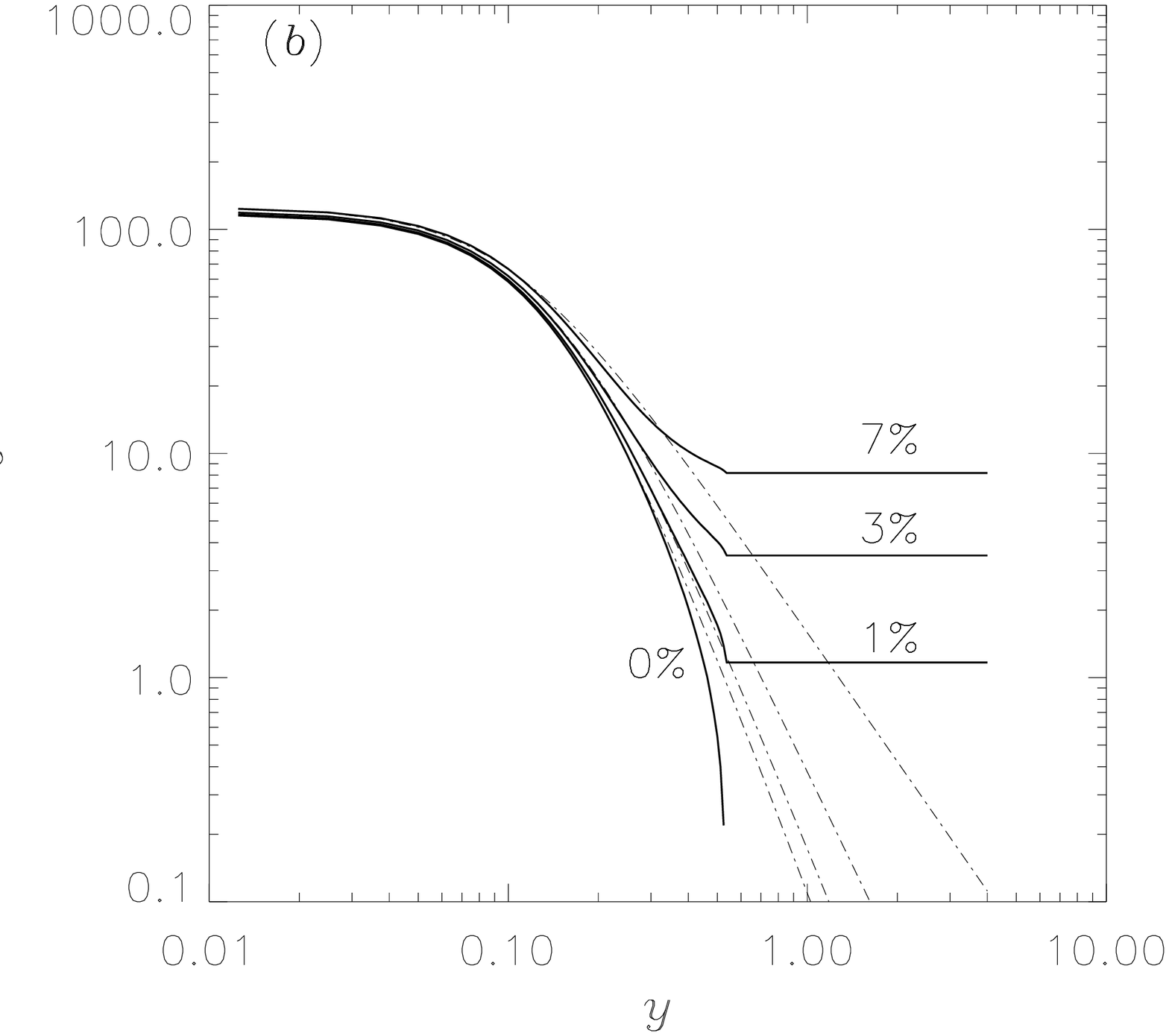}\\
\caption{The column density distribution plotted against the distance
 from the center.
In panel ({\it a}) the column density distribution
 along the $x$-axis, $\sigma(x)=\int \rho(x,y)dy$, is plotted by solid lines 
and
 that along the $y$-axis, $\sigma(y)=\int \rho(x,y)dx$,
 is also shown in ({\it b}).
Background column density is \revII{artificially} added to the magnetohydrostatic 
 solution.
The additional column density is chosen as 0\%, 1\%, 3\%, and 7\% of the maximum
 column densities, $\sigma(x=0)$ and $\sigma(y=0)$. 
Plummer-like distributions to fit the numerical result are plotted
 for respective additional backgrounds by dash-dotted lines (parameter $p$
 is given in Table \ref{table2}).  
\label{fig10}
}
\end{figure}

In Figure \ref{fig10}, the column density integrated along the direction 
 of the $y$-axis, $\sigma(x)=\int\rho dy$,  and that from
 the $x$-axis,  $\sigma(y)=\int\rho dx$, are plotted respectively
 in panels ({\it a}) and ({\it b}), in which  
 the column density is calculated for
 the state with $\rho_c=300$ of Model C3 (Fig.\ref{fig5}{\it c}).
To mimic this kind of observation,
 an additional background column density is added to the magnetohydrostatic
 solution.
We assume the additional column density of 0\%, 1\%, 3\%, and 7\% of
 the column density observed at the center, that is, $\sigma(x=0)$ 
 and $\sigma(y=0)$.
Four curves in this figure correspond to these different backgrounds.
To estimate the parameters,  
 we fit the distribution at three radii:
 at $r=0$, from $\sigma(x=0)$ and $\sigma(y=0)$ we calculate
 the column density at the center, $A\rho_cR_f\equiv \sigma_0$, in equation (\ref{eqn:plummer-sigma});
 we calculate $r=r_{1/2}$ at which the equation gives a half of the
 central column density $\sigma(r_{1/2})=\sigma_0/2$;
 we also calculate $r=r_{1/10}$ at which the equation gives one-tenth  of the
 central column density, $\sigma(r_{1/10})=\sigma_0/10$.
The values of $\sigma_0$, $r_{1/2}$, and $r_{1/10}$ for this model are shown
 in Table \ref{table2}.
The parameter $r_{1/2}$ represents the size of the core and the
 parameter $r_{1/10}$ is related to the steepness of the density distribution
 in the envelope.
Parameters to fit the numerical solution by the Plummer-like distribution
 are also shown in Table \ref{table2}.
Fitted Plummer-like distributions are also shown by dashed-dotted lines
 in Figure \ref{fig10}.
Although the $\chi^2$-fitting might be more appropriate than this
 three-point fitting, Figure \ref{fig10} shows that our fitting works well.

This figure shows that the parameter $p$ takes a value $p \gtrsim 4$
 when the additional background is low ($\lesssim 3\%$) or no background is added (0\%).
This means that the filaments have $p\sim 4$ distribution,
 even if the magnetic field is taken into account.
On the other hand, if we add relatively large background column density, $7\%$
 of $\sigma_0$, a shallow slope in the envelope appears
 and the slope parameter is as small as $p\lesssim 3$.
This means that the slope parameter $p$ is highly dependent
 on the completeness of background subtraction from an observation.
When the background subtraction is incomplete,   
 a shallow slope in the envelope and a small $p$ parameter are expected,
 even if the observed power $p \simeq 2$ may have other origin.

\section{Summary}

We calculated magnetohydrostatic configurations of the isothermal filaments
 that are threaded by the magnetic field laterally.
The magnetic field has an effect in supporting the filament,
 unless the radius of the parent filament $R_0$ is small.
The maximum line-mass supported against the self-gravity  
 is obtained by a function of the magnetic flux threading the filament per unit length
 and the isothermal sound speed.
When considering a filamentary cloud, we have to take the magnetic field
 into account.
  
\acknowledgments

This work was supported in part by JSPS 
 Grant-in-Aid for Scientific Research (A) 21244021 in FY 2009--2012. 
Numerical computations were in part carried out on Cray XT4 and Cray XC30 at 
 the Center for Computational Astrophysics, CfCA, of the National Astronomical 
 Observatory of Japan.

\clearpage

\begin{table}
\begin{tabular}{llllll}
Model & $R_{0}$ & $\beta_0$ & $\rho_c$ range & $\Phi_{\rm cl}$ & $\lambda_{\rm max}$\\
\hline\hline
A\ldots\ldots & 0.5 & 0.03 & $2-10^3$ & 2.89 & 31.6$^*$\\
B1\ldots\ldots & 1   & 0.1  & $2-10^3$ & 3.16 & 33.6$^*$ \\
B2\ldots\ldots & 1   & 0.01 & $2-10^3$ & 10.0 & 60.6$^\times$ \\
C1\ldots\ldots & 2   & 5 & $2-50$ & &\\
C2\ldots\ldots & 2   & 2 & $2-50$ & &\\
C3\ldots\ldots & 2   & 1 & $2-300$ & 2.00 & 28.4$^*$\\
C4\ldots\ldots & 2   & 0.5 & $2-300$ & 2.83 & 31.8$^*$\\
C5\ldots\ldots & 2   & 0.1 & $2-10^3$ & 6.32 & 48.0$^*$\\
C6\ldots\ldots & 2   & 0.01 & $2-10^3$ & 20.0 & 105$^\times$ \\
D1\ldots\ldots & 5   & 1 & $2-300$ & 5.00 & 41.6$^*$ \\
D2\ldots\ldots & 5   & 0.1 & $2-10^3$ & 15.8 & 92.2$^*$ \\
D3\ldots\ldots & 5   & 0.01 & $2-10^3$ & 50.0 & 234$^*$\\
\end{tabular}
\caption{Model parameters and maximum supported mass. \label{table1}
Marks $^*$ and $^\times$ attached to the maximum supported mass 
$\lambda_{\rm max}$ indicate the certainty of respective values and
corresponds to the symbols plotted in Figure \ref{fig9} (see $\S$4.1).}
\end{table}
\clearpage

\begin{table}
\begin{tabular}{llccccc}
Background &       &  $\sigma_0$ & $r_{1/2}$ &  $r_{1/10}$ & $R_f$ & $p$ \\
\hline\hline
0\% \ldots\ldots & $\sigma(x)$ &  66.6 & 0.176 & 0.406  & 0.269 &  4.87\\
0\% \ldots\ldots & $\sigma(y)$ & 117   & 0.0999& 0.235  & 0.146 &  4.6\\
1\% \ldots\ldots & $\sigma(x)$ &  67.2 & 0.178 & 0.421  & 0.256 & 4.53\\
1\% \ldots\ldots & $\sigma(y)$ & 118   & 0.101 & 0.243  & 0.14  & 4.3\\
3\% \ldots\ldots & $\sigma(x)$ &  68.6 & 0.181 & 0.454  & 0.233  & 3.93\\
3\% \ldots\ldots & $\sigma(y)$ & 120   & 0.103 & 0.263  & 0.128   & 3.79\\
7\% \ldots\ldots & $\sigma(x)$ &  71.2 & 0.187 & 0.564  & 0.186  & 2.98\\
7\% \ldots\ldots & $\sigma(y)$ & 125   & 0.106   & 0.327  & 0.103 & 2.92\\
\end{tabular}
\caption{Parameters fitted by Plummer-like distributions for the state
 with $\rho_c=300$ of Model C3. \label{table2}}
\end{table}
\clearpage

\clearpage

\clearpage

\clearpage

\clearpage

\clearpage

\clearpage

\clearpage

\clearpage

\clearpage

\end{document}